\documentclass{aa}  
\usepackage{xcolor}
\usepackage{graphicx}
\usepackage{txfonts}
\usepackage{amsmath}
\usepackage{footnote}\makesavenoteenv{table}
\newcommand{\dbqt}[1]{``#1''}
\def\hii{\mbox{H\,{\sc ii}}}
\def\kms{\mbox{~km\,s$^{-1}$}}

\begin{document}

   \title{Star Formation in Extreme Environments: A 200 pc High Velocity Gas Stream in the Galactic Centre}
   \titlerunning{The 200 pc Helix Stream in the Galactic Centre}


   \author{V. S. Veena\inst{1,2} \and
          W.-J. Kim\inst{2} \and
          {\'A}lvaro S{\'a}nchez-Monge\inst{3,4} \and
          P. Schilke\inst{2} \and
          K. M. Menten\inst{1} \and
          G. A. Fuller\inst{2,5}\and 
          M. C. Sormani\inst{6} \and          
          F. Wyrowski\inst{1} \and
          W. E. Banda-Barrag\'an\inst{7,8} \and
          D. Riquelme\inst{9,10} \and
          P. Tarr\'io\inst{11,12}\and
          P. de Vicente\inst{12}
}
   \institute{Max-Planck-Institut f\"ur Radioastronomie, auf dem H\"ugel 69, 53121 Bonn, Germany\\
              \email{vadamattom@mpifr-bonn.mpg.de}
    \and 
    I. Physikalisches Institut, Universit\"at zu Köln, Z\"ulpicher Str. 77, 50937 K\"oln, Germany
    \and
    Institut de Ci\`encies de l'Espai (ICE, CSIC), Carrer de Can Magrans s/n, E-08193, Bellaterra, Barcelona, Spain
    \and
    Institut d'Estudis Espacials de Catalunya (IEEC), Barcelona, Spain
    \and
    Jodrell Bank Centre for Astrophysics, Department of Physics and Astronomy, The University of Manchester, Manchester M13 9PL, UK    
    \and
    Universit{\`a} dell’Insubria, via Valleggio 11, 22100 Como, Italy
    \and
    Escuela de Ciencias F\'isicas y Nanotecnolog\'ia, Universidad Yachay Tech, Hacienda San Jos\'e S/N, 100119 Urcuqu\'i, Ecuador
    \and
    Hamburger Sternwarte, University of Hamburg, Gojenbergsweg 112, 21029 Hamburg, Germany
    \and
    Departamento de Astronom\'ia, Universidad de La Serena, Av. Cisternas 1200, La Serena, Chile
    \and
    Instituto Multidisciplinario de Investigaci\'on y Postgrado, Universidad de La Serena, Ra\'ul Bitr\'an 1305, La Serena, Chile
    \and 
    Centro de Desarrollos Tecnol\'ogicos, Observatorio de Yebes (IGN), 19141 Yebes, Guadalajara, Spain
    \and
    Observatorio Astron\'omico Nacional (OAN-IGN), 28014, Madrid, Spain
 }
   \date{Received ; accepted }

 
  \abstract
   {The expanding molecular ring (EMR) manifests itself as a parallelogram in the position-velocity diagram of spectral line emission from the Central Molecular Zone (CMZ) surrounding the Galacic centre (GC). It is a high velocity ($\mid V_\textrm{LSR}\mid~>100$\kms) extended molecular gas structure. 
   The formation of the EMR is believed to be associated with the bar driven inflow onto the nuclear region of the Galaxy.  The physical and chemical properties, as well as the evolution of the EMR and its connection to other GC clouds and the CMZ 
   as a whole, are not yet fully comprehended.}
   {Using multiwavelength data, we investigate the gas kinematics, star formation activity,  and the presence of shocked gas in a 200 pc long high velocity gas stream (V$_\textrm{LSR}\sim$+150\kms) with a double helix morphology named the \textit{helix stream}, that is located 15--55 pc above the CMZ ($l\sim0^\circ-1.5^\circ$; $b\sim0.05^\circ-0.4^\circ$) and is kinematically associated with the EMR/parallelogram. }
   {To study the kinematics of the helix stream, we used $^{13}$CO ($J=2-1$) data from the SEDIGISM survey and $^{12}$CO ($J=1-0$) archival data from the Nobeyama telescope.  Additional multiwavelength archival data from infrared to radio wavelengths are used to investigate the star formation activity. We  carried out molecular line observations using the IRAM 30m, Yebes 40m, and APEX 12m telescopes. The detection of four rotational transitions of the SiO molecule ($J=$ 1--0, 2--1, 5--4, 7--6) indicate the presence of shocks. We derived the SiO column densities and abundances in different regions of the helix stream using the rotational diagram method.  We also performed non-local thermodynamic equilibrium (non-LTE) modelling of the SiO emission to analyse the excitation conditions of the shocked gas. }
   {The presence of clumps with submillimeter continuum emission from dust and 
   a candidate \hii~region signify the ongoing star formation activity within the helix stream. The cloud is massive ($2.5\times10^6$~M$_\odot$) and highly turbulent ($\Delta\textrm{V}_\textrm{mean}=18$\kms).  We find evidence of cloud-cloud collisions towards the eastern edge ($l\sim1.3^\circ$), suggesting a dynamic interaction with the CMZ. An expanding shell is detected within the cloud with radius of 6.7~pc and an expansion velocity of 35\kms. The shell might be powered by several supernovae or a single hypernova. The relative abundance of SiO within the helix stream with respect to H$_2$ implies extensive shock processes occurring on large scales ($X$(SiO) $\sim 10^{-9}$). The helical or cork-screw velocity structure observed within the individual strands of the helix stream indicates twisting and turning motions occurring within the cloud.}
   {We propose that the helix stream is the continuation of the near side bar lane, that is overshooting after ``brushing'' the CMZ and interacting with it at the location of the G1.3 cloud. This interpretation finds support both from numerical simulations and prior observational studies of the CMZ. Our findings carry profound implications for understanding star formation in extreme conditions and they elucidate the intricate properties of gas and dust associated with nuclear inflows in barred spiral galaxies.}


   \maketitle
%

\section{Introduction}

\begin{figure*}
\centering
\includegraphics[scale=0.58]{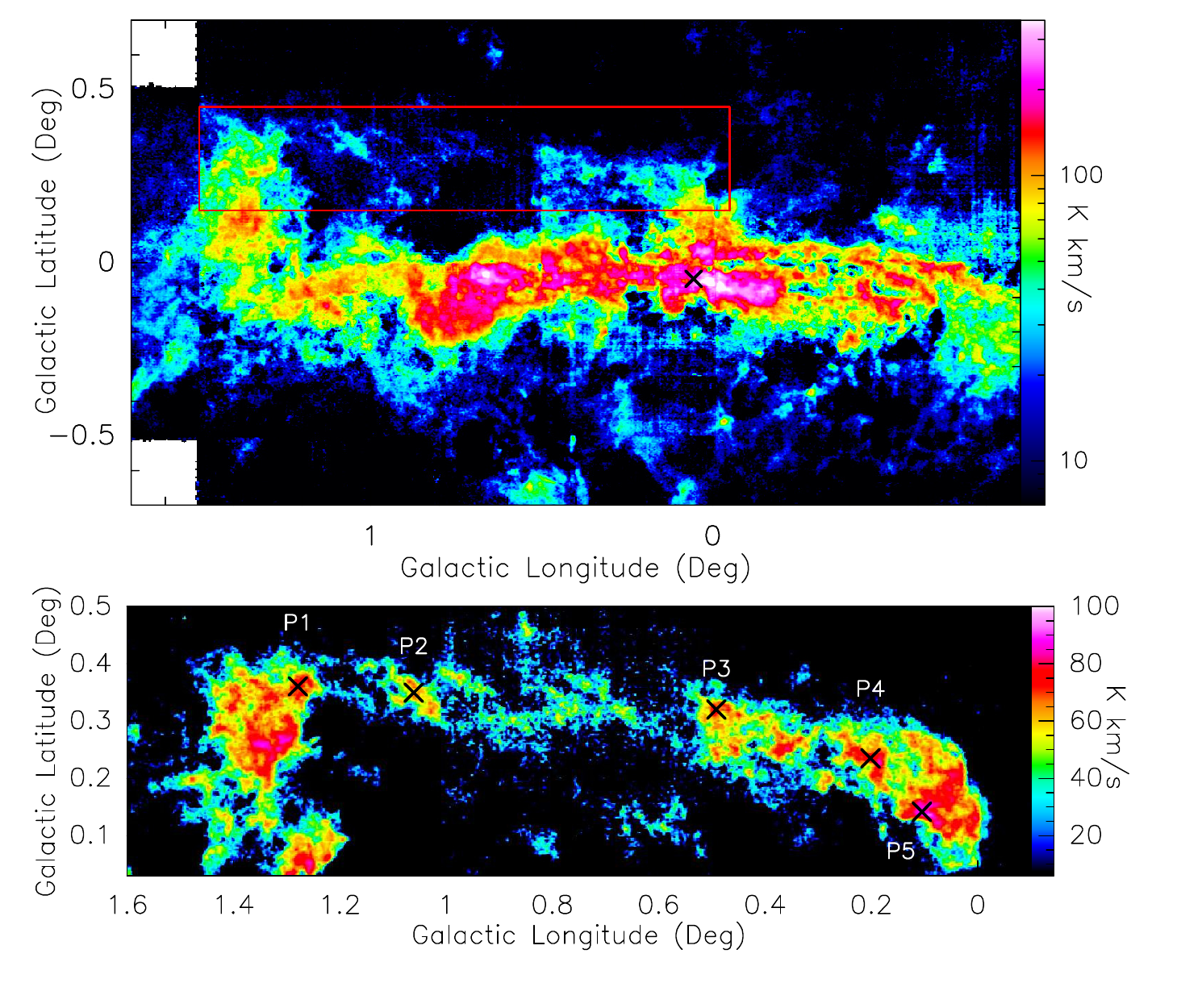} 
\caption{(Top) $^{13}$CO~(2--1) integrated intensity map in the velocity range $-200$ to +200\kms. The box in red indicates the area containing the high velocity stream (right panel) and $\times$ marks the 
positions of Sgr A$^*$.  (Bottom) $^{13}$CO~(2--1) integrated intensity map of the high velocity stream in the velocity range +100 to +200\kms. The five positions used for SiO observations are marked as $\times$ and are labelled.} 
\label{13co}
\end{figure*}

The central part of the Milky Way, the Galactic centre region (GC), located at a distance of 8.2 kpc \citep{{2019ApJ...885..131R},{2019A&A...625L..10G}}, provides an exceptional astrophysical setting to explore the characteristics of radiation and matter within galactic nuclei. It contains the compact non-thermal radio source Sgr A$^*$ that is coincident with the Galactic super-massive black hole \citep{2023A&A...670A..36Y}. The GC hosts a variety of phenomena such as star formation, large-scale gas flows, magnetic loops, and galactic winds \citep{{2018MNRAS.476.5629K},{2021A&A...646A..66P},{2023A&A...674L..15V},{2024MNRAS.530..235Y}}. The inner 1 kpc of the GC harbours a large molecular complex that extends $\sim$400 pc and is known as the central molecular zone \citep[CMZ,][]{1996ARA&A..34..645M}. The molecular gas in the CMZ exhibits complex structures and kinematics as well as peculiar velocity features. This region contains the largest concentration of high-density molecular gas in the Galaxy \citep{2007A&A...467..611F}. The morphology, kinematics, and physical properties and the chemistry of the molecular gas in the CMZ have been investigated in great detail \citep[e.g.,][]{{1987ApJS...65...13B},{1996ARA&A..34..645M},{2012MNRAS.419.2961J},{2013MNRAS.433..221J},{2016A&A...586A..50G},{2016A&A...585A.105L},{2020ApJS..249...35B},{2023ASPC..534...83H}}. 

The mass accretion onto the Galactic nucleus is primarily governed by inflows from kiloparsec scales \citep[e.g.,][]{1999MNRAS.304..512E}. In barred galaxies, the central bar generates a non-axisymmetric gravitational field inducing non-circular gas motions, with gravitational torques crucial for driving mass inflows \citep{{1976Ap&SS..43..491S},{1991MNRAS.252..210B},{2021ApJ...922...79H}}. \dbqt{Dust lanes} are structures associated with galactic bars and can be observed in external galaxies seen face-on \citep{{1961hag..book.....S},{2004ApJ...608..189D},{2023A&A...676A.113S}}. Theoretical models, combined with observational evidence, indicate that these dust lanes serve as pathways through which gas streams onto the CMZ from distances of $\sim$3 kpc \citep[][and references therein]{2019MNRAS.484.1213S}. Identifying these structures, including accretion sites within the CMZ, is challenging due to the solar system's location within a Galactic spiral arm. Classic characterisation methods involve position-velocity (PV) diagrams \citep[e.g.,][]{2008A&A...477L..21M}.  
\begin{figure*}[t]
\centering
\includegraphics[scale=0.55]{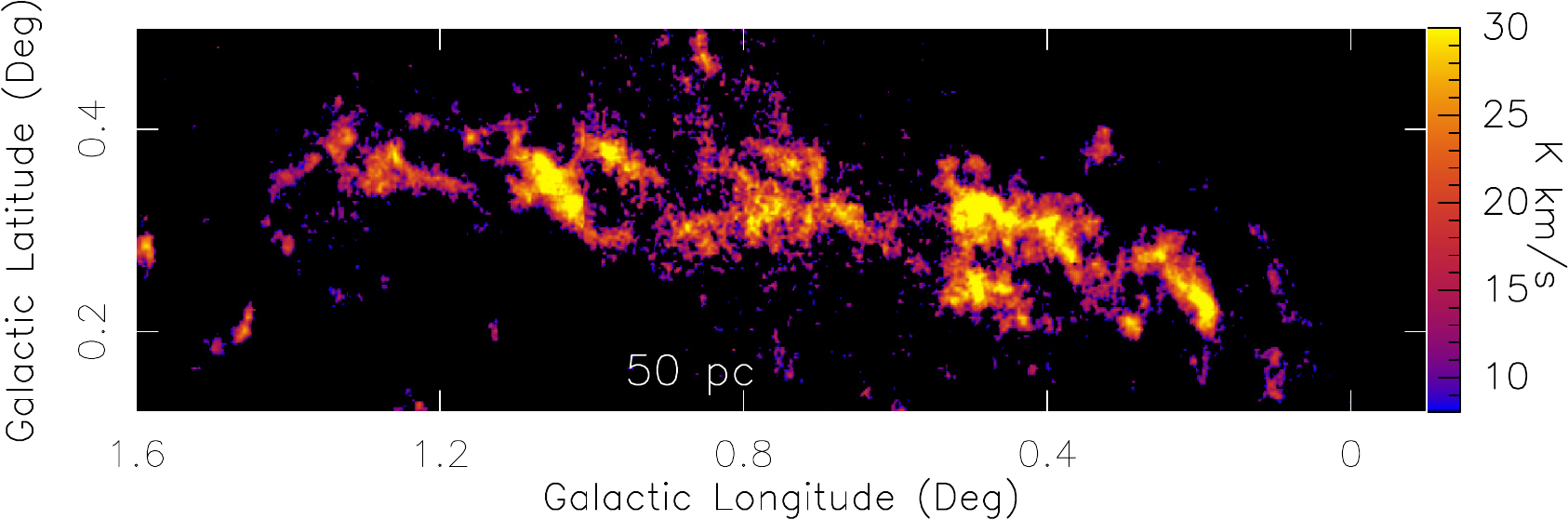} 
\caption{$^{13}$CO~(2--1) intensity map in the velocity range +130 to +190\kms showing the double helix morphology of the helix stream.}
\label{helix}
\end{figure*}

There exists an extended high velocity gas structure in the GC ($\mid l \mid\leq\pm2.0^\circ$), known as the expanding molecular ring (EMR), that is vertically extended to $\sim b\pm0.6^\circ$ \citep{{1972ApJ...175L.127S},{1977ARA&A..15..295O},{1987ApJS...65...13B},{1998ApJS..118..455O}}. The EMR is recognized in PV diagrams as a collection of high-velocity arcs resembling a parallelogram at $\mid V_\textrm{LSR}\mid~>100$\kms (Sofue 2017) and was initially thought to be a radially expanding ring of molecular gas. However, the modern and more widely accepted interpretation is that the EMR has its origins in the non-circular motions driven by the Galactic bar that cause  infalling gas from dust lanes to overshoot the CMZ \citep[see][ and references therein]{2023ASPC..534...83H}.  Although the EMR and the CMZ appear to be located close to each other in the GC, the study of the star formation activity and the physical and chemical properties of the EMR have been limited by the fact that many observational studies of the GC region do not extend enough in latitude or do not have the high sensitivity needed to detect low density emission from the EMR, as its molecular lines have a brightness that is an order of magnitude lower than lines in the CMZ. 

In this work, we investigate the kinematics and star formation activity of the helix stream (Section 3.1) that is spatially and kinematically associated with the EMR. For our analysis, we use multiwavelength data ranging from infrared to radio wavelengths. The organization of the paper is as follows: The details of data are given in Section 2. Section 3 describes the results of our multiwavelength analysis, whereas in Section 4 we discuss our findings. In Section 5 we present our conclusions.


\section{Data}

\subsection{SEDIGISM survey}
The SEDIGISM (Structure, Excitation, and Dynamics of the Inner Galactic Interstellar Medium) project is a spectral line survey of the Milky Way in the $J=2-1$ rotational transition of the  $^{13}$CO and C$^{18}$O isotopologues of the carbon monoxide molecule (CO), which have rest frequencies of 220.398684 and 219.560354 GHz, respectively \citep{2017A&A...601A.124S}. It  was carried out with the Atacama Pathfinder Experiment (APEX) 12-metre submillimetre telescope \citep{2006A&A...454L..13G} and covers 78~deg$^2$ in longitude, $l$, of the inner region of the Galaxy ($-60^\circ<l<+18^\circ$ with a longitude, $b$, coverage of  $|b|<0.5^\circ$). The survey has a larger latitude coverage in the GC region ($|b|<1.0^\circ$). The effective  spectral and angular resolutions are of 0.25\kms and 28$\arcsec$, respectively. The typical root mean square (rms) noise of the survey is 0.8~K in main beam brightness temperature, $T_\textrm{MB}$. In order to study the $^{13}$CO emission towards the GC region, we focused on data within the longitude range $l: 0^\circ-1.5^\circ$, $|b|<1.0^\circ$. We smoothed the original data cube to a velocity resolution of 1\kms for our analysis. The resultant rms noise is 0.4 K.

\subsection{SiO observations with IRAM 30~m telescope}
In order to investigate the presence of shocks within the cloud,  we use the SiO $J$=2--1 line with a rest frequency of 86.84699~GHz. Observations in the 3 mm band were carried out using the Institut de Radioastronomie Millim\'etrique (IRAM) 30-metre telescope during the period 2021 September 09--11 (Project ID: 057-21) with the Eight MIxer Receiver \citep[EMIR;][]{2012A&A...538A..89C}. The standard position switching mode observations towards 5 positions along the cloud (see Fig.~\ref{13co}(Bottom) and Table~\ref{pos}) were covered the frequency range 85.7--109.2 GHz.  The half power beam width (HPBW) of the IRAM telescope is 29$''$ at this 87 GHz.  We used the FTS200 backend with a spectral resolution of 200 kHz (equivalent to 0.6\kms at 3 mm) and, as for the Yebes and Yebes data (see below), the data reduction was performed using the Continuum and Line Analysis Single-dish Software (CLASS) which is part of the GILDAS software \citep{pety2005_gildas}. In order to convert antenna temperatures ($T^*_{\rm{A}}$) to main beam temperatures $T_\textrm{MB}$, we used the relation $T_\textrm{MB}=F_{\rm{eff}}/B_{\rm{eff}} \times T^*_{\rm{A}}$, where $F_{\rm{eff}}=0.95$ is the forward efficiency and $B_{\rm{eff}}=0.81$ is the main beam efficiency. 

\subsection{Yebes 40 m SiO observations}
In order to probe the $J$=1--0 transition of SiO ($\nu_0$=43.42385~GHz),  observations in 7 mm were carried out using the Yebes 40 m telescope on September 27,  2021 (Project ID: 21A017) with the HEMT (High Electron Mobility Transistors) receiver.  It allows broad-band observations with an instantaneous bandwidth of 18 GHz in two linear polarisations. The backends used are Fast Fourier Spectrometers (FFTs) with 2.5 GHz bandwidths and a spectral resolution of 38 kHz \citep{2021A&A...645A..37T}. The HPBW of the Yebes telescope is 40.6$''$ at this frequency. The position switching mode observations towards 5 positions (Fig.~\ref{13co}(Bottom) and Table~\ref{pos}) covered the frequency range 31.5--50 GHz. 
The spectra were measured in units of antenna temperature and in order to convert this to T$_\textrm{MB}$, we used a beam efficiency factor of 0.55.

\subsection{APEX SiO observations}
To probe the $J$=5--4 ($\nu_0$=217.10492~GHz) and  $J$=7--6 ($\nu_0$=303.9268~GHz) transitions of SiO, we used the APEX telescope (Project IDs: M9518C$\_$109, M9518A$\_$109). The position switching observations were carried out towards 5 positions (Fig.~\ref{13co}(Bottom) and Table~\ref{pos}) using the nFLASH230 receiver and the 7 pixel Large APEX sub-Millimetre Array (LASMA). The 2 sideband nFLASH230 receiver is connected to two FFT backends, providing a simultaneous coverage of two sidebands with a bandwidth of 7.9 GHz each. The separation between the centre of the two sidebands is 16 GHz. LASMA pixels consists of sideband separating mixers (2SB) which provides an IF bandwidth of 4--8 GHz for each of the two sidebands. The HPBW are 28.5$''$ and 20$''$ for the SiO~(5--4) and (7--6) lines, respectively.  In order to convert antenna temperatures to T$_\textrm{MB}$, we used beam efficiency factors of 0.83 and 0.74 for the nFLASH230 and LASMA receivers.

\section{Results}

\subsection{A high velocity molecular cloud in the GC with double helix morphology}

Fig.~\ref{13co}(Top) shows the $^{13}$CO~(2--1) integrated intensity map of the GC region in the velocity range from $-200$ to +200\kms. The image shows emission that extends predominantly along, mostly slightly below, the Galactic plane, which represents the CMZ. On the positive latitude end of the distribution, diffuse emission is observed that arches toward higher latitudes, up to $\sim 0.4^\circ$. 
Fig.~\ref{13co}(Bottom) presents the zoomed-in view of the above mentioned region, integrated within the smaller velocity range +100 to +200\kms. Bright, elongated emission extending over $\sim1.4^\circ$ (in longitude) is observed. The width of the emission feature is $\sim0.15^\circ$ (in latitude) and the corresponding aspect ratio is 9.3, implying that the structure is filament-like in morphology. It has a tilt with an inclination angle $\sim8.5^\circ$, with respect to the Galactic plane. The CO cloud exhibits multiple clumpy substructures. The high velocity of the cloud ($>$100\kms) cannot be explained with a simple Galactic rotation model \citep{2017MNRAS.470.1982S}. Its unique location in the GC and large velocities suggest its association with the EMR. Assuming that the cloud is at the distance of the GC (D=8.2$\pm1$~kpc), we estimate its extent  to be 200~pc~$\times ~21.5$~pc. The cloud is located $15-55$~pc above the Galactic mid-plane.


\begin{figure}[t]
\centering
\hspace*{-0.2cm}
\includegraphics[scale=0.32]{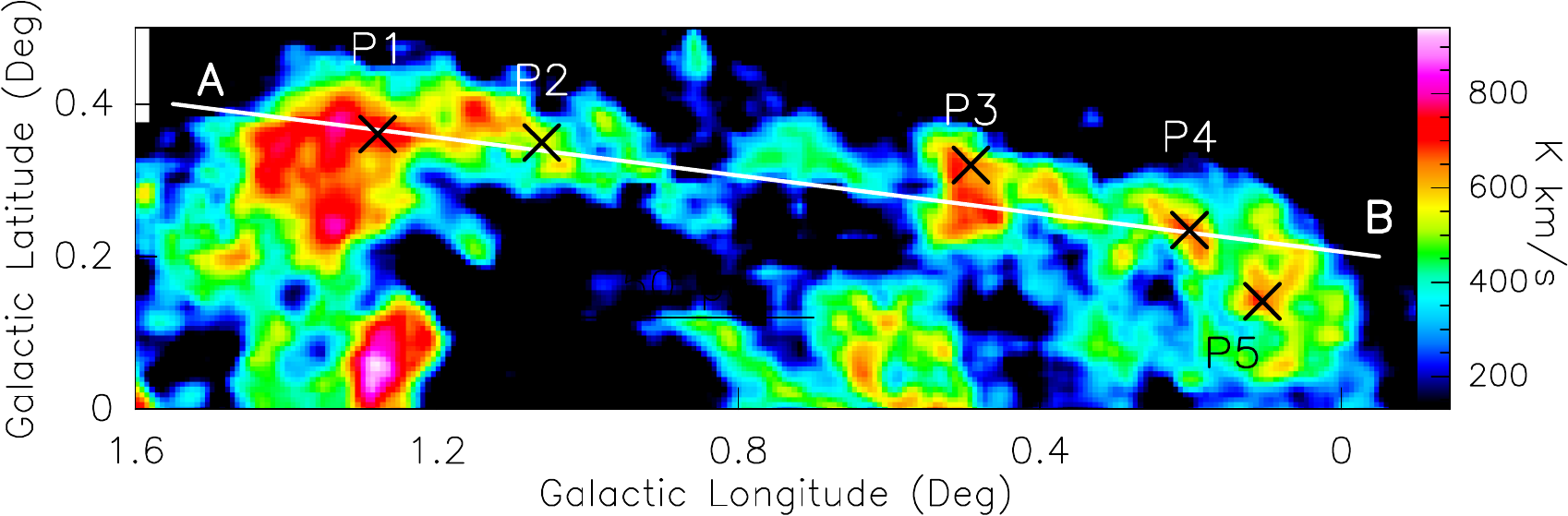} 
\caption{$^{12}$CO(1--0) integrated intensity map \citep{1998ApJS..118..455O} of the helix stream (shown in Fig.~\ref{13co}(Right) and Fig.~\ref{helix}) in the velocity range 100--200\kms. The white straight line AB used to construct the PV diagram is marked. The five positions used for SiO observations are marked as $\times$ and are labelled.} 
\label{12co}
\end{figure}
\begin{table}
\centering
\caption{List of 5 positions used for spectral line observations with IRAM, APEX and Yebes telescopes}
\begin{tabular}{c c c c c}
\hline\\
Position&RA(J2000) &Dec(J2000)&$l$&$b$\\
&($^{h\,m\,s}$)&($^\circ$ $'$ $''$)&(Deg)&(Deg)\\
\hline\\
P1&17:47:14.23&--27:39:23.22 &1.278&0.361\\
P2&17:46:46.15&--27:50:54.03 &1.061&0.349\\
P3&17:45:32.49&--28:20:59.21 &0.492&0.320\\
P4&17:45:11.09&--28:38:30.31 &0.202&0.235\\
P5&17:45:19.04&--28:46:22.86 &0.105&0.142\\
\hline
\end{tabular}
\label{pos}
\end{table}

To investigate further the intricate substructure of the cloud, we created nine intensity maps integrated within 10\kms~velocity intervals, which  are presented in Fig.~\ref{channel}. Several clumpy features are observed. The lowest velocity features correspond to the emission from the edges of the filament (100--130\kms). The emission from the central region is evident in higher velocity images (+130 to +190\kms). In order to analyse the morphology of the emission from the central region, which can be visualised better with a lower dynamic range image, we generated an additional integrated intensity map of emission in the high velocity range (i. e., +130 to +190\kms). The resultant image is presented in Fig.~\ref{helix}. The emission has a remarkable \dbqt{double helix} morphology containing two helically wound strands. The origin of this double helix morphology is explored in Section 4.3. The cloud will be addressed as the \textit{helix stream} in the following sections.

\begin{figure*}[!htbp]
\centering
\includegraphics[scale=0.8]{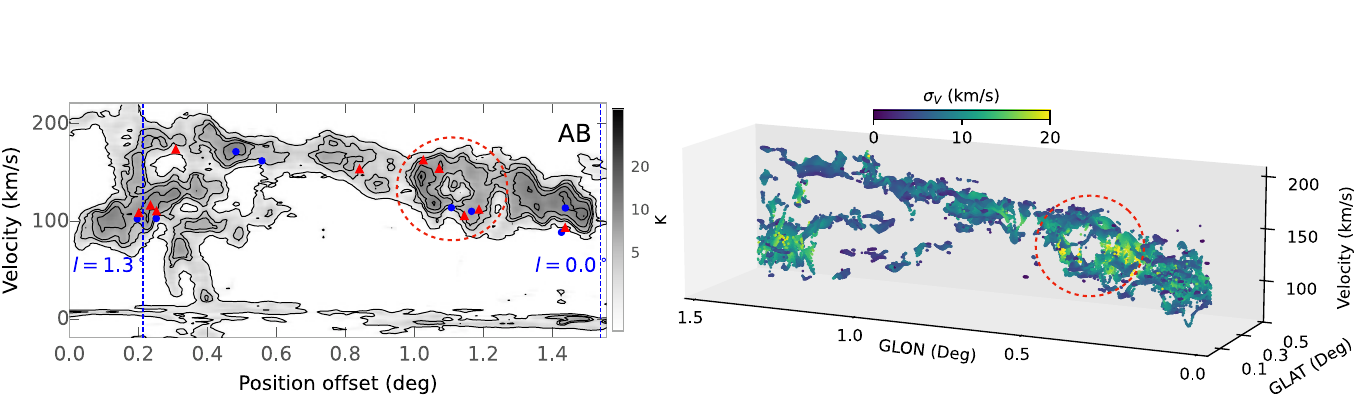}
\caption{(Left) $^{12}$CO(1--0) PV diagram of the helix stream along the cut AB indicated  in Fig.~\ref{12co} overlaid with the position-velocity loci of Hi-GAL prestellar (blue dots) and protostellar (red triangles) clumps discussed in Section 3.4. (Right) The $^{13}$CO~(2--1) PPV plot of the helix stream generated from spectral decomposition using $\tt{GAUSSPY+}$. Each velocity component is colour coded according to its velocity dispersion. Ellipse marks the shell feature discussed in Section 4.1.} 
\label{pv}
\end{figure*}

\subsection{Kinematics of the helix stream}

\subsubsection{Position-velocity diagram}
In order to study the velocity structure and large-scale kinematics of the helix,  we used $^{12}$CO maps from the GC CO survey by \citet{1998ApJS..118..455O}, who mapped  the GC region in the $^{12}$CO ($J$=1--0) transition using the 45~m telescope at Nobeyama Radio Observatory.  The spectral resolution and grid spacing are 0.65\kms and 34$\arcsec$, respectively. The advantage of using $^{12}$CO (1--0) emission (critical density $\approx2\times10^3$~cm$^{-3}$) to study the large-scale cloud kinematics is that we are sensitive to regions of low density (and low column density) and thus can also probe 
the velocity structure of the diffuse envelope surrounding the helix stream. However, it is important to note that in denser (and higher column density) regions, opacity effects may occur, potentially affecting the velocity measurements. Fig.~\ref{12co} shows the integrated intensity map of the $^{12}$CO emission integrated over the same velocity range as in Fig.~\ref{13co}(Right). The overall morphology and spatial extent of the $^{12}$CO(1--0) emission distribution is similar to that of the $^{13}$CO~(2--1) emission except for the fact that the  $^{12}$CO emission appears more extended and clumpier compared to the sharp emission features observed in the latter, only tracing the denser regions. To explore the velocity distribution of the region, we construct a PV diagram along the major axis of the cloud (see Fig.~\ref{12co}). 

 From the PV diagram (Fig.~\ref{pv}(Left)), we identify a wide variety of emission features with velocities ranging from $\sim$0 to +200\kms. The emission corresponding to the helix stream is seen as an elongated feature in the velocity range $\sim100-200$\kms. Towards the eastern edge of the cloud ($l\sim1.3^\circ$), several clumpy features are observed in the  velocity range from $\sim 70$ to 100\kms. These are connected to the high velocity emission with bridge-like features. Such bridge-like features have been found to exist in several Galactic molecular clouds and are believed to be signatures of cloud-cloud collisions \citep[e.g.,][]{{Haworth2015},{2019MNRAS.488.4663S}}.  The bridge feature corresponds to the region in the 2D intensity map (Fig~.\ref{13co}(Right)) where the filament slightly arches downwards towards the CMZ. As the 100\kms clumps are further connected to lower velocity features consistent with velocities of CMZ clouds (0--100\kms), this indicate a possible interaction between the CMZ and the helix stream. Another prominent feature in the PV diagram is a ring-like structure extending $\sim0.26^\circ$ (36 pc), centred at $l\sim0.42^\circ$ (marked as a dashed ellipse in Fig.~\ref{pv}(Left)).  Such ring or arc-like features are associated with expanding shells of gas within the cloud \citep{2011ApJ...742..105A} and are the  signatures of possible stellar feedback. We explore the origin of the expanding shell in Section 4.1.  Apart from velocity sub-structures such as the bridge, the ring and the 100\kms clump, a large scale monotonous velocity variation of $\sim100$\kms along the filament is seen, which 
 corresponds to a gradient of 0.5\kms\,pc$^{-1}$.  

\subsubsection{Spectral decomposition using $\tt{GAUSSPY+}$}

To analyse the velocity distribution of the dense gas in detail, we employed pixel-by-pixel spectral decomposition of the $^{13}$CO~(2--1) data using the $\tt{GAUSSPY+}$ package \citep{{2015AJ....149..138L},{2019A&A...628A..78R}}.  We chose data of the $^{13}$CO~(2--1) line instead of $^{12}$CO(1--0) as $^{13}$CO is  less abundant than $^{12}$CO by factors of $\sim$20--30 in the GC, and its lines are less affected by optical depth effects. $\tt{GAUSSPY+}$ is based on $\tt{GaussPy}$, a python based Autonomous Gaussian Decomposition algorithm, which is used to analyse complex spectra arising in  the interstellar medium (ISM) by decomposing them into multiple Gaussian components.  It involves a technique called derivative spectroscopy, and automatically determines the initial guesses for the Gaussian components for individual spectra.  In the later stages of the fitting routine,  $\tt{GAUSSPY+}$ also carries out an automated spatial refitting based on neighbouring fit solutions.  We adopted the default parameters for the spectral decomposition provided by \citet{2019A&A...628A..78R} except for the smoothing parameters $\alpha1$ and $\alpha2$ for which  we generated a training set consisting of 5000 random spectra from the data cube.  Using the training set, we estimated  $\alpha1$ and $\alpha2$ to be 1.23 and 4.55, respectively. The spectra towards the bridge region were fitted with up to 8 components. However, on average, three components were fitted to a typical spectrum.

In Fig.~\ref{pv}(Right), all the identified components above a signal-to noise ratio of 5 and velocities exceeding 70\kms are presented as a position-position-velocity (PPV) plot. The data are colour coded according to the velocity dispersion of the corresponding component. The overall morphology of the features observed in the PPV diagram is consistent with that of the $^{12}$CO PV diagram, though the extended diffuse emission is not as evident as in the $^{12}$CO PV diagram owing to the fact that the $^{13}$CO emission traces higher column density gas compared to the former. The bridge feature seen in the $^{12}$CO PV diagram is also observed in the PPV plot. The ring-like feature identified in the PV diagram is clearly evident in the $^{13}$CO PPV plot as a spherical shell with a central cavity. 

\begin{figure}[t]
\centering
 \includegraphics[scale=0.47]{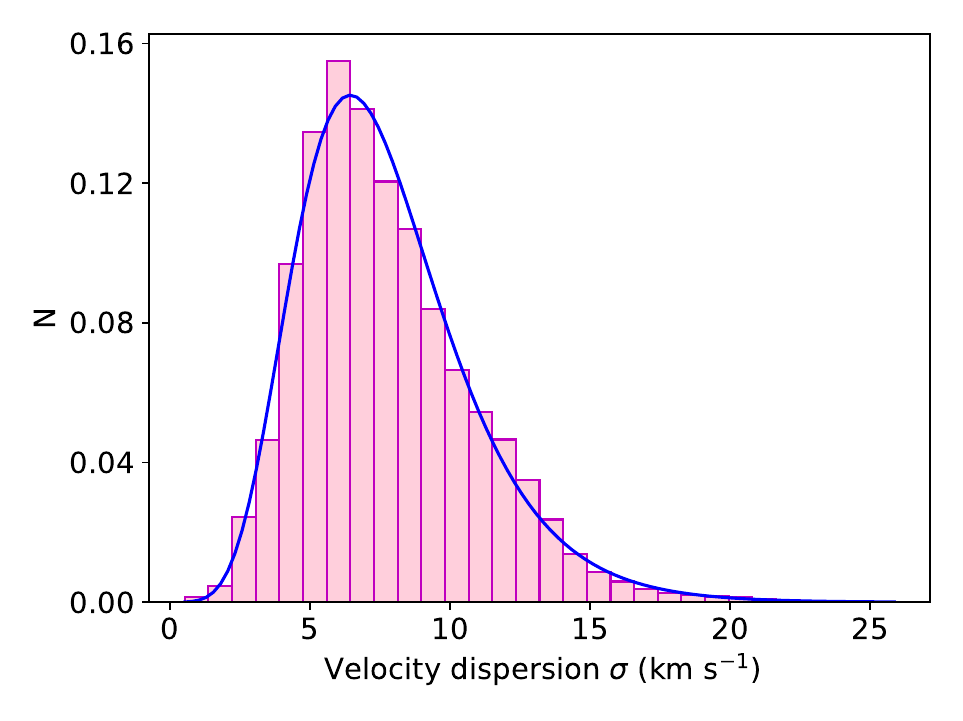}
\caption{Histogram of velocity dispersion of spectra decomposed using $\tt{GAUSSPY+}$. Blue line corresponds to the lognormal fit to the histogram.} 
\label{disphist}
\end{figure}

Fig.~\ref{disphist} presents a histogram of velocity dispersions (based on Full Width at Half Maximum (FWHM) of the $^{13}$CO emission line) from the spectral decomposition using {\tt{GAUSSPY+}}. The velocity dispersion is found to range from 1.0 to 25\kms with a mean of 7.8\kms ($\sigma_\textrm{median}=$~7.2\kms). As seen in the plot, the histogram of the velocity dispersion follows a lognormal distribution. The observed velocity dispersions in the helix stream are consistent with velocity dispersion measurements of clouds in the CMZ region \citep[$\sigma_\textrm{median}=$~9.8\kms;][]{2016MNRAS.457.2675H}. Highest velocity dispersion (i.e., $\sigma>15$\kms) is observed mainly towards the 100\kms eastern clump, the connecting bridge, and the shell-feature at $l\sim0.42^\circ$. As the observed line width could have contributions from thermal as well as non-thermal motions, we proceed to estimate the magnitudes of each component using the following expressions \citep[e.g.,][]{1983ApJ...270..105M} 

\begin{equation}
\sigma_\textrm{T} = \sqrt{\frac{k_\mathrm{B} T_\mathrm{kin}}{\mu m_\mathrm{H}}}  
\end{equation}
\begin{equation}
\sigma_\textrm{NT} = \sqrt{(\sigma)^2-(\sigma_\textrm{T})^2},  
\end{equation}
\noindent where $\sigma_\textrm{T}$ and $\sigma$ correspond to the thermal and the observed dispersion, $T_\textrm{kin}$ is the kinetic temperature, $\mu$ is the molecular weight of $^{13}$CO that is 29, $k_\textrm{B}$ is the Boltzmann constant and $m_\textrm{H}$, the mass of the hydrogen atom. Following the approach of \citet{2016MNRAS.457.2675H}, we assume a fiducial temperature range of 60--100 K typical for the CMZ. 
This corresponds to thermal velocity dispersion ranging between 0.1--0.2\kms and is thus mostly neglible. 
The obtained non-thermal $\sigma_\textrm{NT}$ dispersion can be used to estimate the Mach number using the expression
\begin{equation}
    \mathcal{M}_{3D} \approx \sqrt{3} \frac{\sigma_\textrm{NT}}{c_\mathrm{s}}
\end{equation}

\noindent where $c_s=\sqrt{k_\textrm{B} T_\textrm{kin}/\mu_p m_\textrm{H}}$ is the isothermal sound speed and $\mu_\textrm{p}$=2.33 is the mean weight for molecular hydrogen gas. For the same temperature range of 60--100~K, $c_\textrm{s}$ is found to range from 0.5--0.6\kms. For the mean velocity dispersion of 7.8\kms, the Mach number $\mathcal{M}_{3D}$ ranges between 23--27 at a scale of 1~pc (equivalent to the angular resolution of the $^{13}$CO data). This suggests that on parsec-scales, the gas motions within the helix stream are on the average, highly supersonic. However, there is a caveat that the obtained Mach number could be an upper limit, in case of ordered gas flows and substructures present at sub-parsec scales that are unresolved.  Towards the circular shell and the eastern bridge seen in the PV and PPV plots, the Mach numbers exceed 50. This could be explained as increased turbulence as a result of expanding gas motions within the cavity and/or cloud-cloud collisions along the bridge. In all cases, our estimates of $\mathcal{M}_{3D}$ suggest that the gas harbours strong turbulence.

\subsubsection{Size-velocity dispersion relation}

Spectral line observations of molecular clouds reveal  a power-law correlation between the velocity dispersion and the projected size, with a power-law index ranging between 0.2 to 0.6 \citep[e.g.,][]{{1981MNRAS.194..809L},{1987ApJ...319..730S},{1995ApJ...446..665C}}. This is known as the Larson's scaling relationship.  In this section, we examine the velocity dispersion-size relationship of the helix stream derived from$^{13}$CO~(2--1) data. For this, we first identify the relevant structures within the cloud using the dendrogram-based structure identification python package $\tt{ASTRODENDRO}$ \citep{2008ApJ...679.1338R}. The dendrogram is a powerful technique to analyse hierarchical structures within a molecular cloud. In order to compute the dendrogram, three parameters are required: (i) min$\_$value, the minimum intensity threshold taken as 3$\sigma_\textrm{rms}$ where $\sigma_\textrm{rms}$ is the rms noise, (ii) min$\_$delta that is the minimum height of the identified structure, taken to be 1.5$\sigma_\textrm{rms}$, and (iii) min$\_$npix, the minimum number of pixels within a structure, considered to be 6, equivalent to two beams in area. The structures are categorised into trunks (largest continuous structures), branches (intermediate structures), or leaves (smallest substructures). For each structure, we compute the size R, extracted from the geometric mean of the semi-major and semi-minor axis (radius parameter in $\tt{ASTRODENDRO}$) i.e., R = $\eta\sqrt{\sigma_{maj}\sigma_{min}}$ where $\eta$, the factor that relates the radius of a spherical cloud to its one-dimensional rms size,  is assumed to be 1.91 \citep[e.g.,][]{{1987ApJ...319..730S},{2021A&A...656A.101M}}. The velocity dispersion $\sigma$ corresponds to the v$_\textrm{rms}$ parameter, that is the intensity-weighted second moment of velocity. Using $\tt{ASTRODENDRO}$, we identify 635 distinct structures within the cloud. 

\begin{figure}[t]

\includegraphics[scale=0.55]{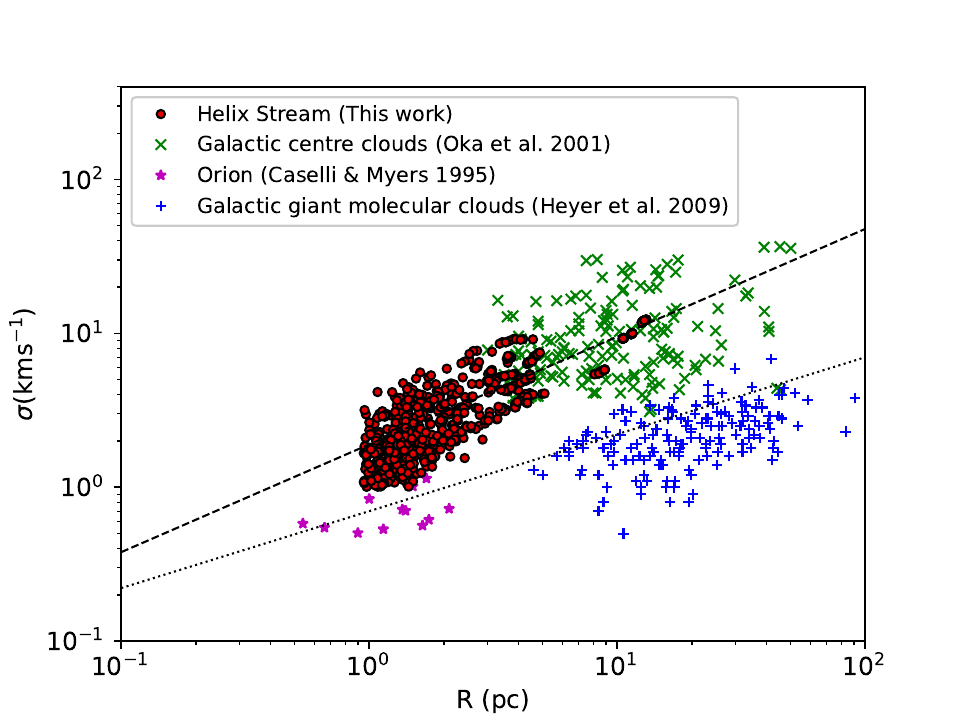} \quad \hspace*{-0.3cm}\includegraphics[scale=0.537]{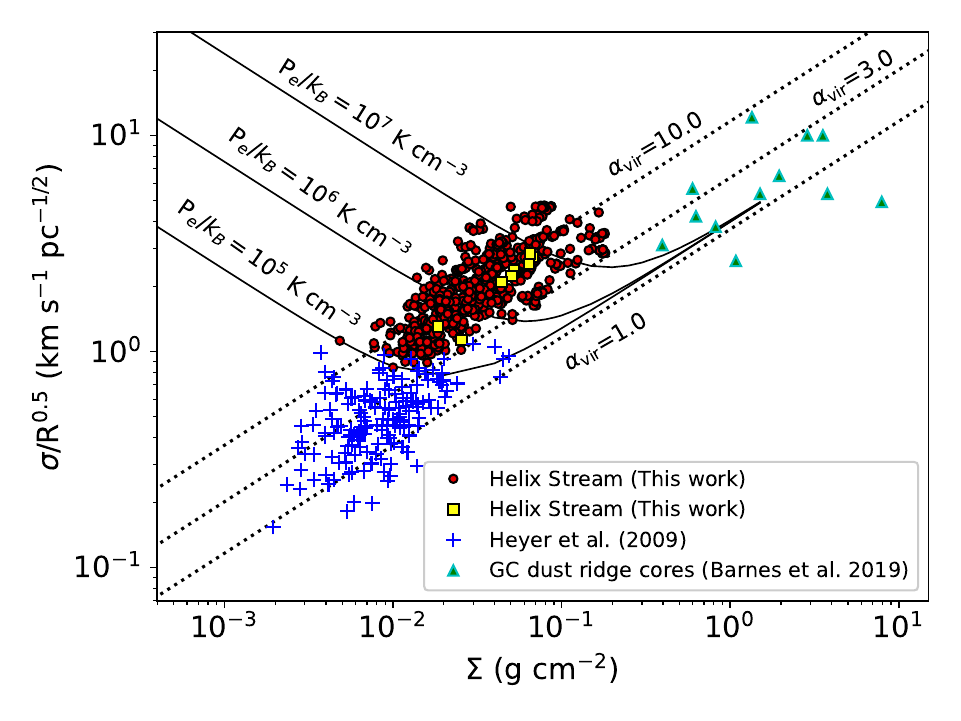}
\caption{(Top) Comparison of size versus velocity dispersion in the helix stream (red dots) with values of Galactic centre clouds from \citet[][green $\times$]{2001ApJ...562..348O}, the Orion A and B  giant molecular clouds \citep[magenta stars;][]{1995ApJ...446..665C} and other Galactic giant molecular clouds \citep[blue crosses;][]{2009ApJ...699.1092H}. The dotted line corresponds to the best-fit result from \citet{1987ApJ...319..730S}, $\sigma=0.7$R$^{0.5}$. The dashed line corresponds to the best-fit result for the helix stream $\sigma=2.97$R$^{0.70}$. (Bottom) Heyer's relation (surface density, $\Sigma$, versus $\sigma$/R$^{0.5}$) for different Galactic clouds. Dots correspond to the structures within the helix stream,  squares correspond to structures within the helix stream where follow-up molecular observations are carried out, crosses represent Galactic giant molecular clouds \citep{2009ApJ...699.1092H}, and triangles represent high-density cores in the GC \lq dust-ridge' clouds \citep{2019MNRAS.486..283B}. Black curves indicate solutions of pressure-bounded virial equilibrium and dotted lines correspond to $\alpha_\textrm{vir}$=1.0, 3.0, and 10.0 where there is negligible external pressure.}
\label{radsig}
\end{figure}
\begin{figure*}[t]
\centering
\includegraphics[scale=0.55]{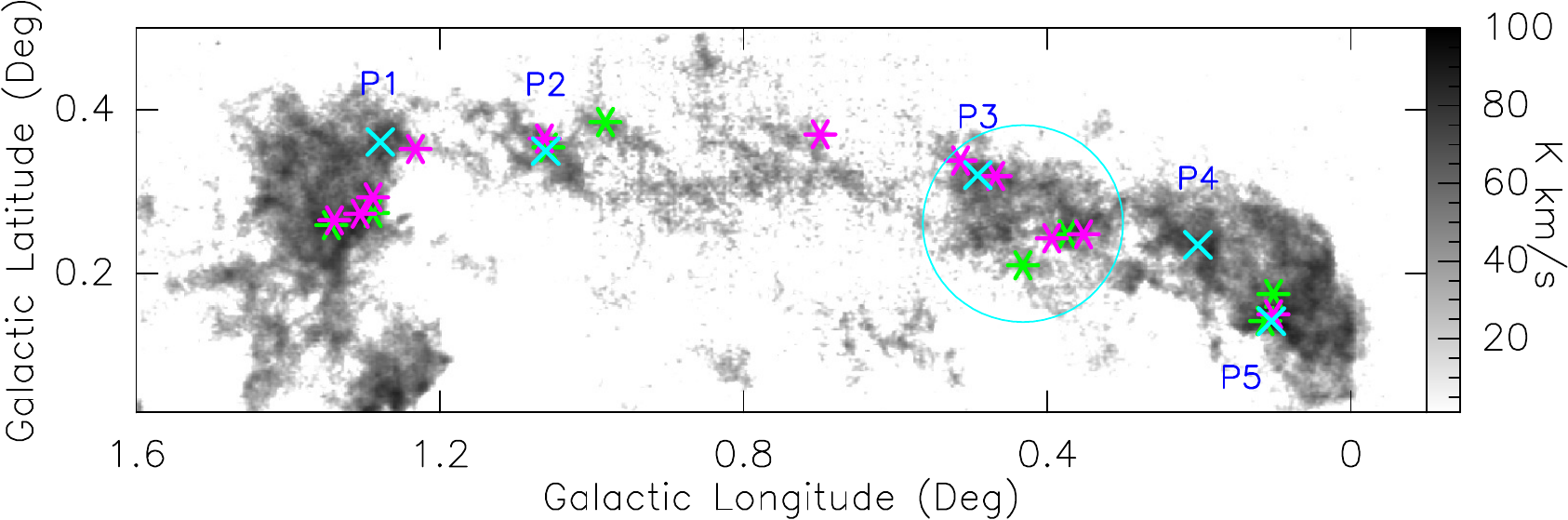}
\caption{$^{13}$CO integrated intensity map overlaid with Hi-GAL prestellar (green asterisks) and protostellar (magenta asterisks) clumps. Red triangle corresponds to the location of the radio source GPSR5 0.431+0.262. Blue circle roughly corresponds to the location of ring-like feature observed in the PV diagram. The five positions used for SiO observations are marked as $\times$.}
\label{hgal_rad}
\end{figure*}

The $\sigma$--R plot of the helix stream is presented in Fig.~\ref{radsig}(Top). The largest structures identified within the cloud have R$\sim$10--20 pc and the corresponding velocity dispersions are in the range of 9--12~\kms, whereas a systematic decrease is observed in the velocity dispersions of smaller structures. For comparison, we have also plotted data corresponding to other Galactic clouds \citep{{1995ApJ...446..665C},{2001ApJ...562..348O},{2009ApJ...699.1092H}}. From the plot, it appears that $\sigma$ is correlated with R. To quantify this, we fitted the data with a power-law of the form
\begin{equation*}
\left(\frac{\sigma}{\kms}\right) = \mathrm \alpha\,\left(\mathrm{\frac{R}{pc}}\right)^\beta    
\end{equation*}
From the $\chi^2$ fit, we estimated the power-law index $\beta$ and coefficient $\alpha$ as $0.7\pm0.01$ and $1.90\pm0.05$, respectively. The obtained index is steeper than the classical power-law index of 0.5 \citep{1981MNRAS.194..809L} and is comparable to that of CMZ clouds \citep[$\sim 0.66$,][]{{2012MNRAS.425..720S},{2017A&A...603A..89K}}.
\begin{table*}[!htbp]
\centering
\caption{Properties of Herschel Hi-GAL compact sources identified within the helix stream}
\begin{tabular}{c c c c c c c c c c c}
\hline\\
Source ID&GLON&GLAT&Diameter&Distance\tablefootmark{a}&Velocity&Mass\tablefootmark{b} &$T_\textrm{dust}\tablefootmark{b}$&$L_\textrm{Bol}$&$\Sigma$&Class\tablefootmark{c}\\
&(Deg)&(Deg)&(pc)&(kpc)&(\kms)&(10$^2$M$_\odot$)&(K)&(10$^3$L$_\odot$)&(g\,cm$^{-2}$)&\\
\hline\\
HIGALBM0.1020+0.1502&0.1020&0.1502&1.9&8.4&93.7&55.5&16.4&14.2&0.39&2\\
HIGALBM0.1017+0.1750&0.1017&0.1749&1.1&8.4&113.8&50.5&14.9&4.4&1.19&1\\
HIGALBM0.1129+0.1423&0.1129&0.1423&1.3&8.4&88.9&14.3&15.9&2.3&0.22&1\\
HIGALBM0.3521+0.2479&0.3521&0.2479&1.0&8.4&112.4&9.3&15.5&2.5&0.27&2\\
HIGALBM0.3736+0.2481&0.3736&0.2481&1.0&8.4&110.4&43.2&12.0&1.1&1.13&1\\
HIGALBM0.3944+0.2428&0.3944&0.2428&1.1&8.4&105.9&14.3&15.0&2.0&0.34&2\\
HIGALBM0.4322+0.2101&0.4322&0.2100&0.9&8.4&114.1&1.5&28.2&7.3&0.05&1\\
HIGALBM0.4677+0.3197&0.4677&0.3197&0.8&8.4&154.2&4.1&16.7&0.9&0.16&2\\
HIGALBM0.5138+0.3377&0.5138&0.3377&1.4&8.3&162.9&9.4&14.8&1.4&0.12&2\\
HIGALBM0.6988+0.3696&0.6988&0.3696&1.4&8.4&153.8&2.2&17.9&0.9&0.03&2\\
HIGALBM0.9817+0.3846&0.9817&0.3846&1.2&8.4&162.1&17.1&13.6&1.0&0.31&1\\
HIGALBM1.0573+0.3538&1.0573&0.3538&0.7&8.4&171.6&55.7&10.4&0.5&3.02&1\\
HIGALBM1.0620+0.3648&1.0620&0.3648&0.4&8.4&166.8&2.5&18.8&1.0&0.48&2\\
HIGALBM1.2317+0.3524&1.2317&0.3524&0.9&8.4&174.1&12.2&14.8&1.5&0.38&2\\
HIGALBM1.2876+0.2928&1.2876&0.2928&1.5&8.6&110.6&33.3&13.0&1.8&0.41&2\\
HIGALBM1.2882+0.2743&1.2882&0.2743&1.0&8.6&103.1&51.4&10.8&0.7&1.47&1\\
HIGALBM1.3052+0.2733&1.3052&0.2733&0.7&8.6&116.6&5.7&14.9&0.7&0.31&2\\
HIGALBM1.3390+0.2649&1.3390&0.2649&1.2&8.6&108.8&21.9&13.0&1.9&0.43&2\\
HIGALBM1.3435+0.2596&1.3435&0.2596&1.2&8.6&102.7&137.4&10.3&1.3&2.64&1\\
\hline
\end{tabular}
\tablefoot{\tablefoottext{a}{\citet{2021A&A...646A..74M}},\tablefoottext{b}{Obtained by fitting the SEDs in the spectral range 160~$\mu$m$\leq\lambda\leq$500~$\mu$m} with a modified blackbody function, \tablefoottext{c}{1: prestellar, 2: protostellar}}
\label{higaltab}
\end{table*}

\subsubsection{Cloud stability and dynamics}
In order to investigate whether the helix stream is stable against gravitational collapse, we examined the modified Larson's relationship, known as the Heyer relationship \citep{2009ApJ...699.1092H}. The surface densities ($\Sigma$) of the dendrogram structures are calculated and plotted against $\sigma$/R$^{0.5}$ in the bottom panel of Fig.~\ref{radsig}. We find a systematic variation of $\sigma$/R$^{0.5}$ with $\Sigma$. The structures within the helix stream deviate from the loci of gravitationally bound clouds where the virial parameter, $\alpha_\textrm{vir}$, equals 1. The stability of a cloud can be defined based on the virial critical parameter ($\alpha_\textrm{crit}$). Unbound, sub-critical clouds are characterised by $\alpha_\textrm{vir} > \alpha_\textrm{crit}$, whereas gravitationally bound, supercritical clouds are characterised by $\alpha_\textrm{vir} < \alpha_\textrm{crit}$. According to \citet{2013ApJ...779..185K}, for non-magnetised clouds, $\alpha_\textrm{crit} \gtrsim 2$, whereas for strongly magnetised clouds, $\alpha_\textrm{crit} \ll 2$. \citet{2009ApJ...699.1092H} estimated the mean $\alpha_\textrm{vir}$ as 1.9, whereas for the helix stream, it ranges between 2.7 and 36.5, with a mean of 9.5, implying that a majority of the structures within the helix stream are highly sub-critical and unbound. \citet{2011MNRAS.416..710F} suggested that the external pressure ($P_\textrm{e}$) introduces an additional confining force that is not accounted under the simple virial equilibrium assumption, which only considers gravitational and kinetic energies. Based on their analysis, the data from \citet{2009ApJ...699.1092H} that appears to systematically deviate from the simple virial equilibrium can be in pressure-bounded virial equilibrium (PVE), if the external force $P_\textrm{e}/k_\textrm{B}$ is in the range 10$^4$--10$^6$~K\,cm$^{-3}$. The solutions to the PVE, as proposed by \citet{2011MNRAS.416..710F}, can be written as

\begin{equation}
\frac{\sigma^2}{R}=\frac{1}{3}(\pi\Gamma\,G\Sigma+\frac{4P_\mathrm{e}}{\Sigma})    
\end{equation}

\noindent where $\Gamma$ is related to the density structure of the cloud. Here, for simplicity, we assume $\Gamma=0.73$, corresponding to a centrally concentrated density structure \citep[e.g.,][]{{2011MNRAS.416..710F},{2018MNRAS.474.2373W}}. The solutions to the PVE appear to be V-shaped in the $\Sigma$ versus $\sigma$/R$^{0.5}$ plot, observed in the bottom panel of  Fig.~\ref{radsig}. Assuming that the helix stream is in PVE, the external pressure has to be of the order of $P_\textrm{e}/k_\textrm{B}\sim10^5-10^7$~K\,cm$^{-3}$. In Section 3.2.2, we find that there is highly supersonic turbulence within the helix stream. This turbulent pressure could be responsible for the PVE within the helix stream.

\begin{figure}[h!tbp]
\centering
\includegraphics[scale=0.5]{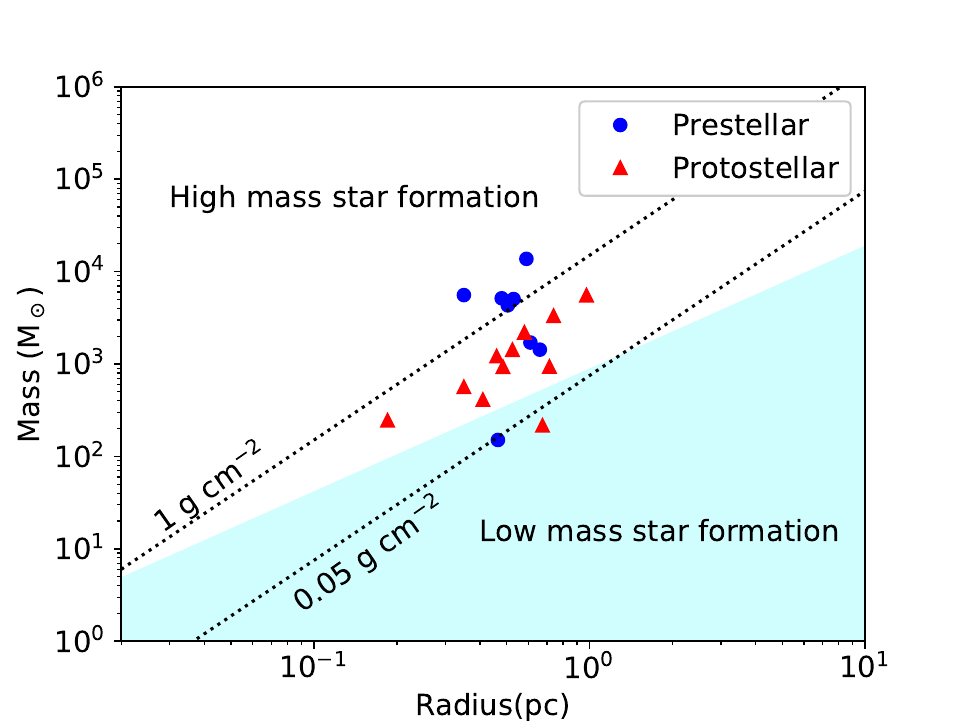}\quad \includegraphics[scale=0.5]{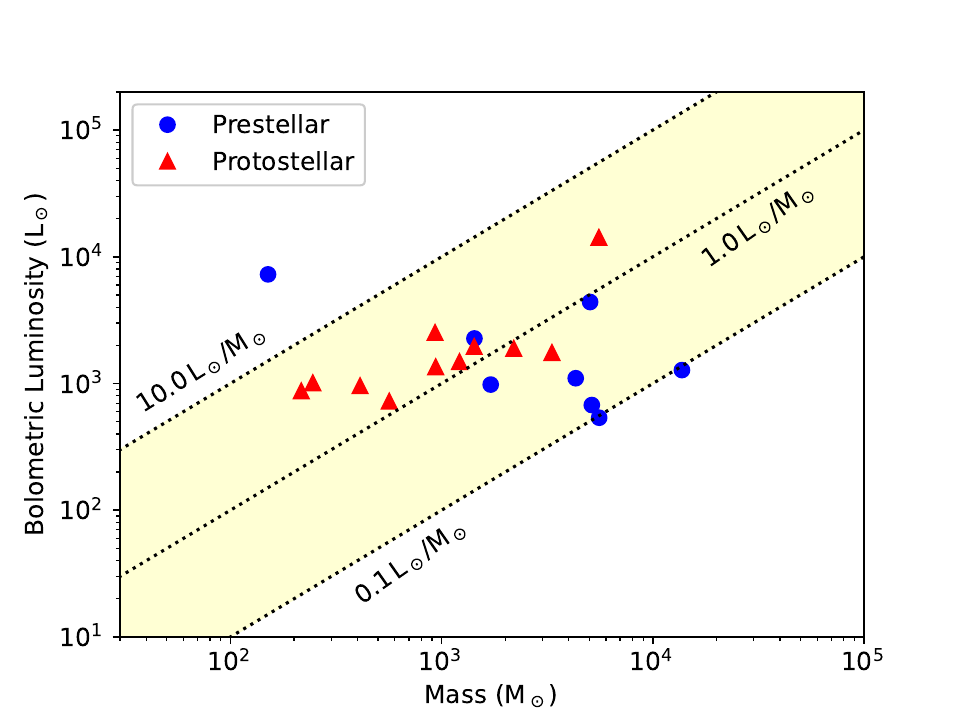}
\caption{(Top) Mass-radius relation of 19 Hi-GAL clumps. The shaded area represents range of masses consistent with low mass star formation, satisfying the criteria M$\leq$870~M$_\odot$~(r/pc)$^{1.33}$ \citep{2010ApJ...723L...7K}. The dotted lines indicate surface density thresholds of 1 and 0.05~g\,cm$^{-2}$, respectively. (Bottom) Bolometric luminosity-mass relation of 19 Hi-GAL clumps. Dotted lines represent ($L/M$)=0.1, 1.0, and 10.0~$L_\odot$/$M_\odot$, respectively.}
\label{hgal}
\end{figure}

\subsection{Column density and mass}

To estimate the mass and column densities of the cloud, we followed the X-factor method, to convert the integrated intensities of $^{13}$CO~(2--1) emission to H$_{2}$ column density. For this, we used the mean conversion factor X$_{^{13}\textrm{CO}(2-1)}$ = 
$1.08\times10^{21}$ (K\kms) estimated by \citet{2017A&A...601A.124S}. The integrated intensity map in the velocity range +100 to +200\kms is converted to a column density map. The peak column density is estimated to be $8.7\times10^{22}$~cm$^{-2}$, whereas the average column density along the filament is $2.0\times10^{22}$~cm$^{-2}$. The total column density N($\textrm{H}_2$) can be used to estimate the mass of the helix stream using the equation 
\begin{equation}
\mathrm{M=N(H_2)\,\mu_g\,m_H\,A}   
\label{cden_mass}
\end{equation}

\noindent where M is the total mass of the cloud, $\mu_g$ is the mean weight of molecular gas taken as 2.86 assuming that the gas is 70$\%$ molecular hydrogen by mass \citep{2010A&A...518L..92W}, m$_\textrm{H}$ is the mass of the hydrogen atom, and A is the area of the cloud calculated by adding the area of individual pixels within the 5$\sigma$ contour of $^{13}$CO emission. The total mass of the helix stream is thus estimated to be $2.5\times10^6$~M$_\odot$. We would like to caution that the estimated mass could be affected by optical depth effects towards the peak positions of the high density clumps as well as a lower X$_{^{13}\textrm{CO}(2-1)}$ factor in the GC region \citep[e.g.,][]{{1998A&A...331..959D},{2023ApJ...959...93G}}.
\begin{figure*}[!htbp]
\centering
\hspace*{-0.3cm}
\includegraphics[scale=0.25]{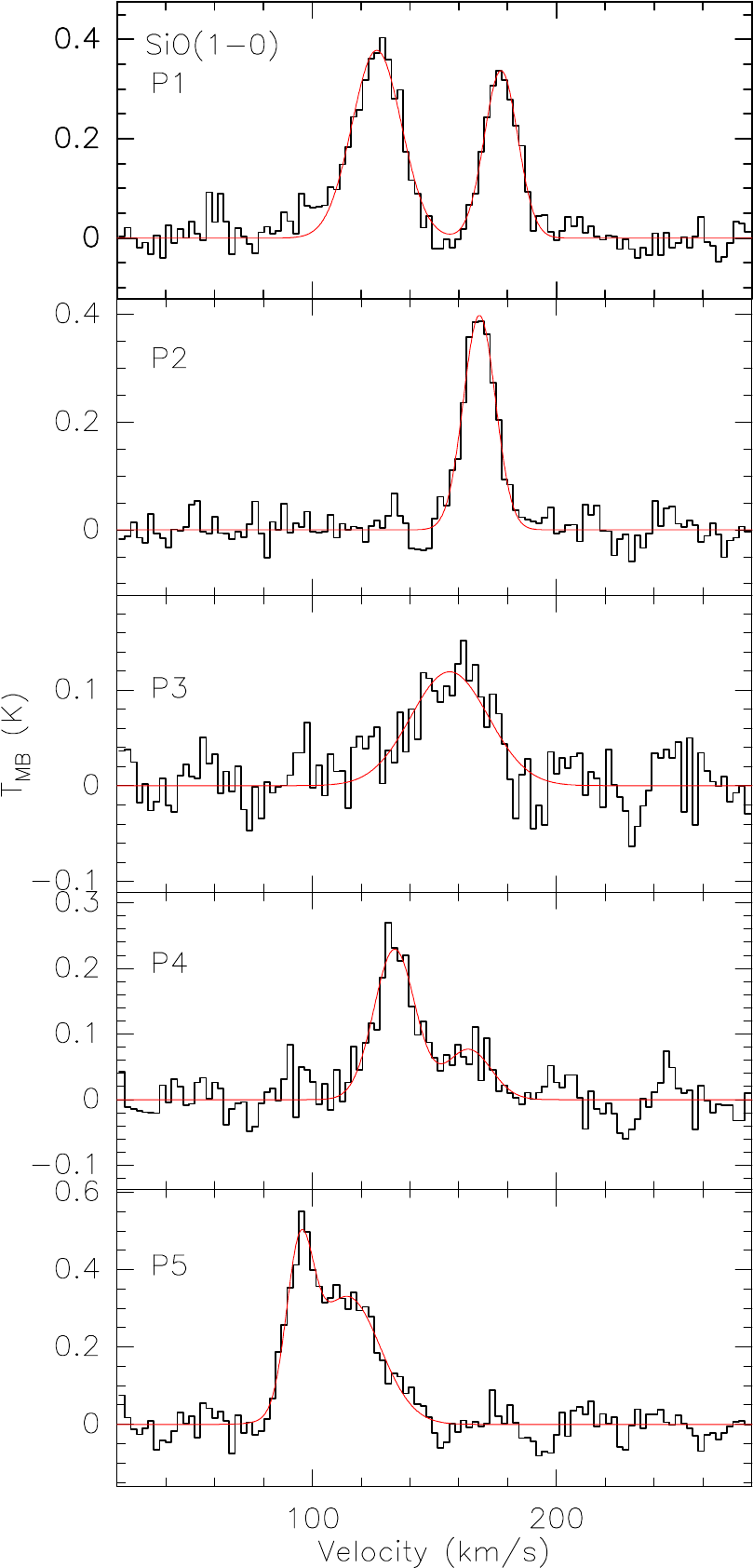}\quad\includegraphics[scale=0.25]{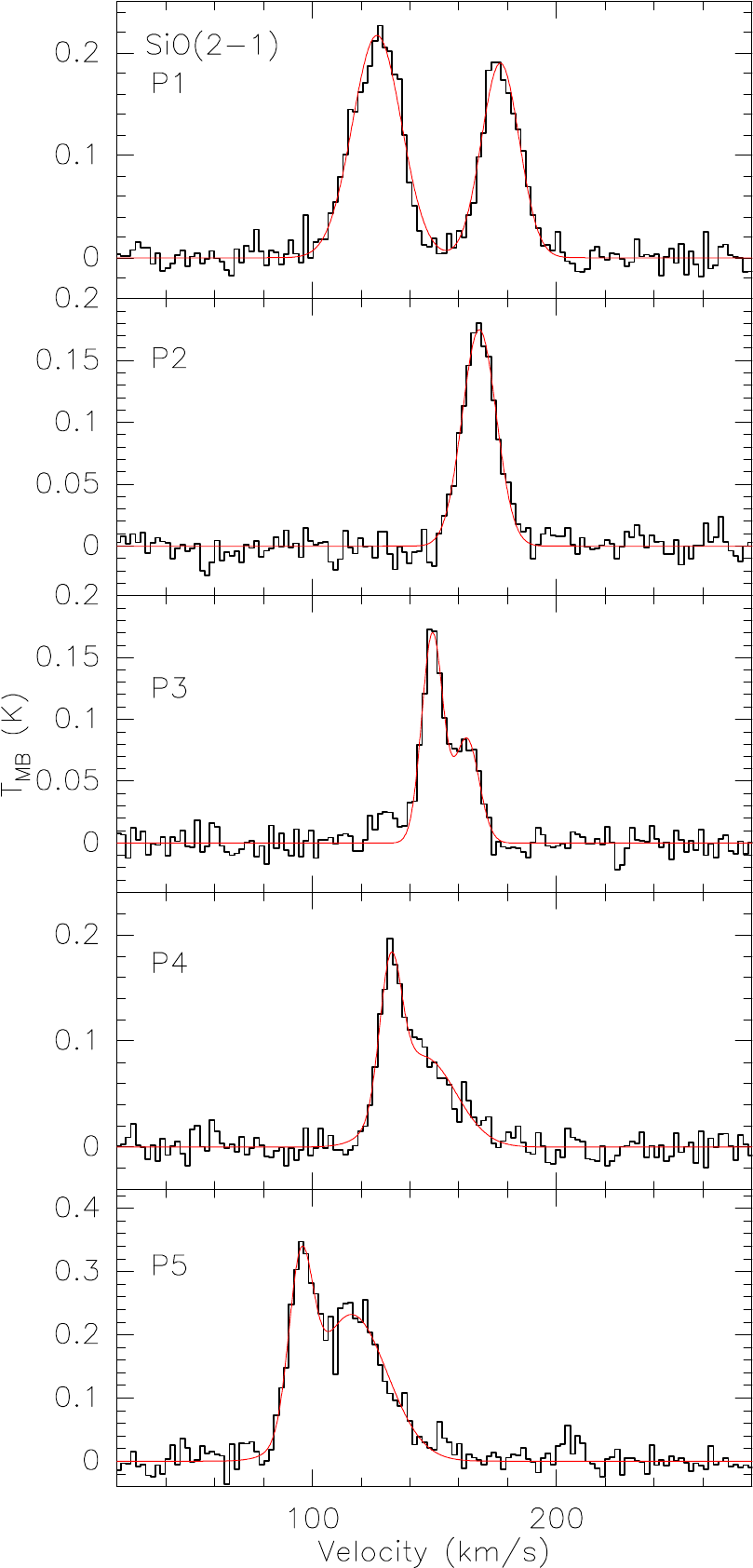}\quad\includegraphics[scale=0.25]{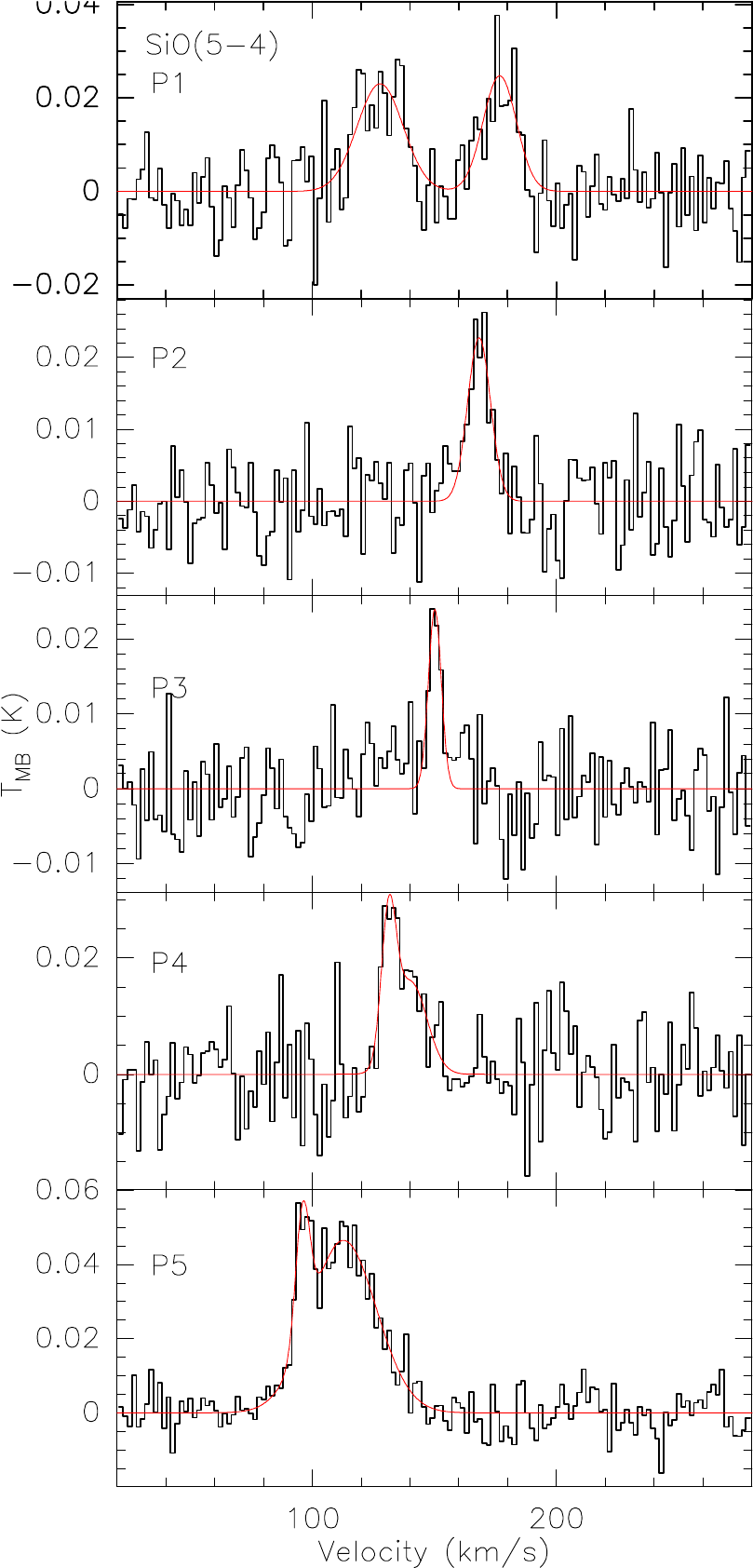} \quad\includegraphics[scale=0.25]{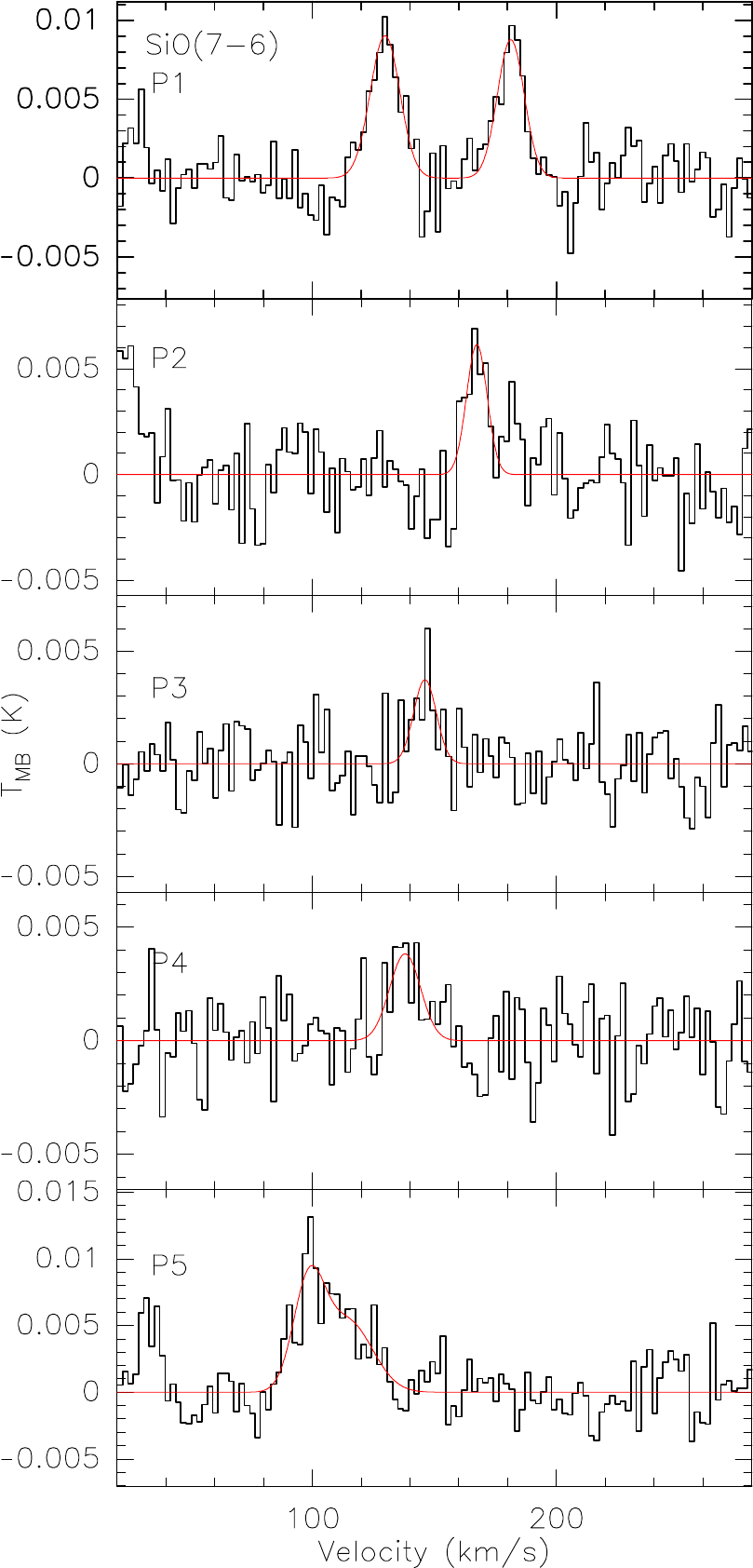}\quad\includegraphics[scale=0.25]{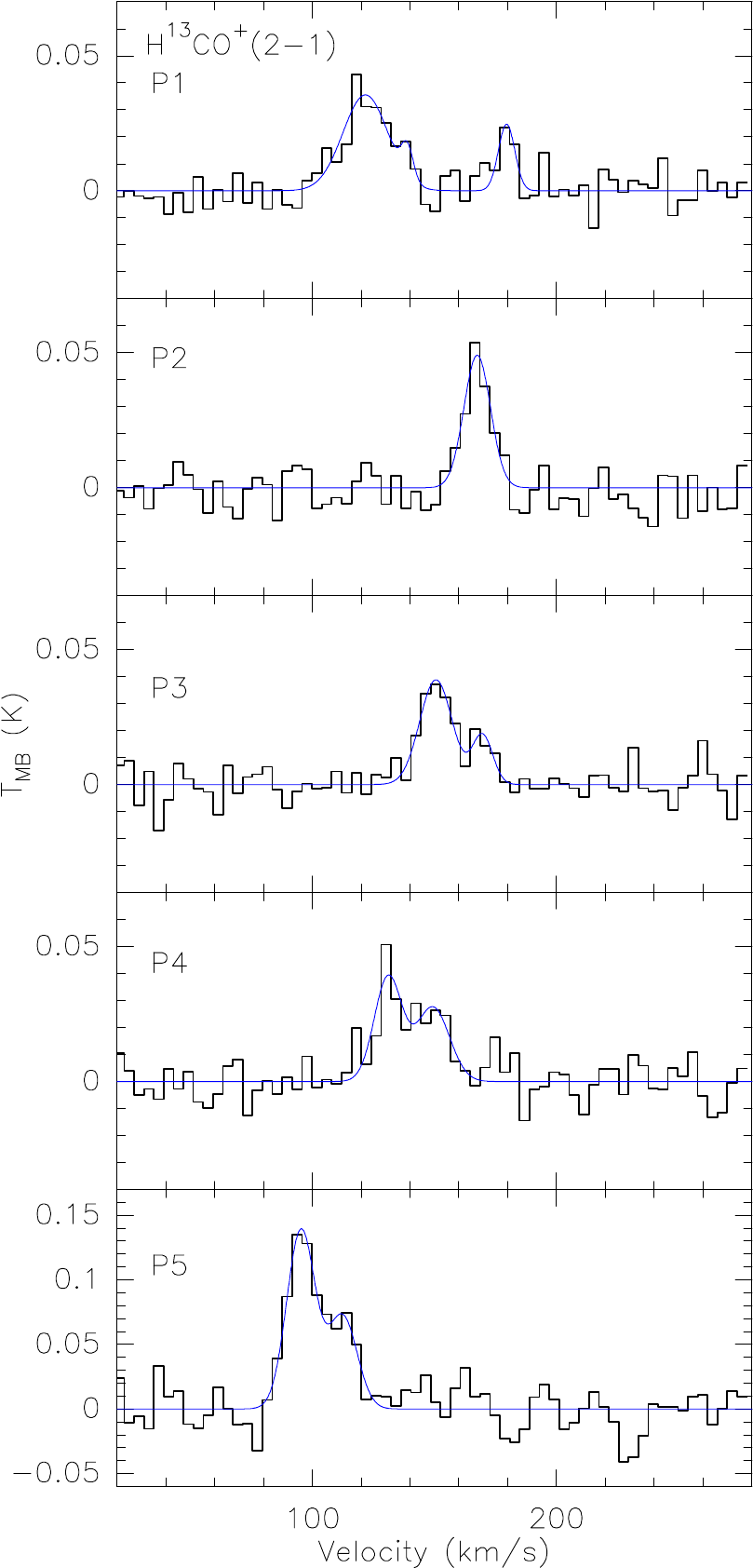}
\caption{SiO (1--0), (2--1), (5--4) and (7--6) and H$^{13}$CO$^+$(2--1) spectra toward positions P1, P2, P3, P4 and P5. Gaussian fits to the SiO and H$^{13}$CO$^+$(2--1) spectra are shown in red and blue, respectively.}
\label{sio}
\end{figure*}
We also estimated column densities and mass of the cloud using the H$_2$ column density map produced by the point process mapping (PPMAP) method \citep{2017MNRAS.471.2730M}. The PPMAP uses Hi-GAL continuum data in the wavelength range 70--500~$\mu$m and it takes complete account of the point spread functions (PSFs) of the telescope used, resulting in a spatial resolution of 12$''$. Dust temperatures and column densities are determined by fitting a spectral energy distribution (SED) assuming an opacity of 0.1~cm$^{2}$g$^{-1}$ at 300~$\mu$m, a power-law index $\beta$ of 2, and a dust to gas ratio of 100. The PPMAP derived dust temperature and column density maps of the helix stream region are presented in Fig.~\ref{Tcd}. The peak column density is $1.2\times10^{23}$~cm$^{-2}$ and the mean column density is $2.3\times10^{22}$~cm$^{-2}$, similar to the column density estimates based on $^{13}$CO emission. We estimate the total mass of the cloud from the column density using equation~\ref{cden_mass}. Using the PPMAP method we find the mass of the cloud to be $4.2\times10^6$~M$_\odot$. This estimate is an upper limit as the source is in the GC region and there could be contributions from dust associated with the ISM along the line of sight. In addition, there could be uncertainties due to the assumed dust to gas ratio. The agreement between the mass estimates from X-factor and PPMAP methods, each falling in the order of $10^6$~M$_\odot$, gives confidence in our results, despite methodological variations and potential uncertainties. 

\subsection{Dust clumps associated with the helix stream}

In order to study the properties of the cold dust clumps associated with the cloud, we use the Herschel Hi-GAL compact source catalogue \citep{{2017MNRAS.471..100E},{2021MNRAS.504.2742E}} which is a band-merged catalogue of cold dust clumps identified in the Herschel wavelength bands (70--500~$\mu$m). The catalogue comprises sources that are detected in at least three consecutive Herschel bands. The distances to the clumps are estimated based on methods described in \citet{2021A&A...646A..74M} and physical properties are computed using a single temperature graybody fit to the SEDs. Clumps are classified into three different categories: starless clumps, prestellar clumps, and protostellar clumps. Starless clumps are objects that are gravitationally unbound. Prestellar clumps are gravitationally bound clumps that are not associated with 70~$\mu$m sources. Protostellar clumps are those clumps with star formation activity, i.e., 
indicated by the presence of 70~$\mu$m sources.

From the Hi-GAL catalogue, we found 19 clumps within the 5$\sigma$ contour of the $^{13}$CO intensity map. We selected only clumps whose velocities range between 70 and 200\kms, corresponding to velocities of observed kinematic features in the PV diagram. The distribution of Hi-GAL clumps along the helix stream is shown in Fig.~\ref{hgal_rad} and properties of individual clumps are listed in Table~\ref{higaltab}. Of 19 clumps, 8 are in the prestellar phase and 11 are in the protostellar phase. Masses of clumps range between $1.5 and 137.4\times10^2$~M$_\odot$ and luminosities range between $0.7$ and $14.2\times10^3$~M$_\odot$. The radius-mass relation of these sources is presented in Fig.~\ref{hgal}(Top). Of 19 clumps, 17 satisfy the criteria for massive star formation defined by \citet{2010ApJ...723L...7K}, i.e.,  M~$>$~870M$_\odot$(r/pc)$^{1.33}$. The surface densities of 5 prestellar clumps exceed the surface density threshold of 1~g\,cm$^{-2}$ for massive star formation, defined by \citet{2008Natur.451.1082K}. In Fig.~\ref{hgal}~(Bottom), we present the mass-luminosity relationship of the Hi-GAL clumps, which is considered to be a good diagnostic of a clump's  evolutionary \citep{2008A&A...481..345M}. The bolometric luminosity to mass (($L/M$)) ratio of all, except for one clump, lies in the range 0.1--10.0~$L_\odot$/$M_\odot$. Of these, 8 clumps have ($L/M$)~$<$~1.0~$L_\odot$/$M_\odot$. Overall, the low ($L/M$) ratio of the Hi-GAL clumps suggests that they are in relatively early evolutionary phases \citep{2017A&A...603A..33G} in which the gas is still accumulated and compressed (($L/M$)~$\lesssim$~2.0~$L_\odot$/$M_\odot$) or the young stellar objects gaining mass (2.0\,$L_\odot$/$M_\odot$~$\lesssim$~($L/M$)~$\lesssim$~40.0~$L_\odot$/$M_\odot$). The distribution of 19 Hi-GAL clumps in the position-velocity space, as depicted in Fig.~\ref{pv}(Left), reveals a notable concentration towards the bridge and shell features in the PV diagram. This spatial clustering suggests a scenario of triggered star formation, wherein the interaction of molecular gas within these regions leads to enhanced densities and the initiation of gravitational collapse. Such triggering could be driven by external factors such as shock fronts, turbulence, or feedback effects.  

\subsection{Massive star formation within the expanding CO shell}

In Section 3.2, we identified ring/shell-like structures in the PV and PPV plots that are associated with a shell-like feature in the integrated intensity map. Here we explore the origin of the CO shell and its implications on the star formation activity in the helix stream. The approximate centre of the cavity in the intensity plot is at ($l,b$)=($0.42^\circ, 0.26^\circ$) with an angular diameter of $\sim9.0\arcmin$,  which translates to a diameter of 22 pc (see Fig.~\ref{hgal_rad}). As mentioned in Section 3.2.1, an expanding shell of gas would create a ring-like feature in the PV and PPV plots \citep{2011ApJ...742..105A}. The velocity of the circular feature in the PV diagram ranges between 100--170\kms, which translates to an expansion velocity of $\sim$35\kms. 

\begin{figure*}[!htbp]
\centering
\includegraphics[scale=0.36
]{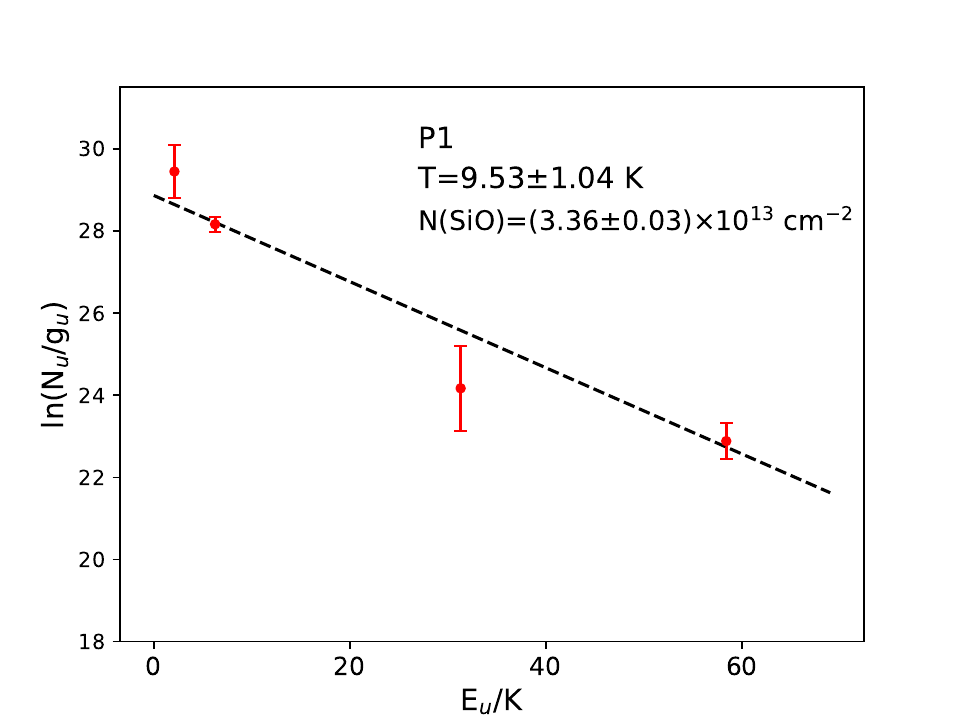}\quad\includegraphics[scale=0.36]{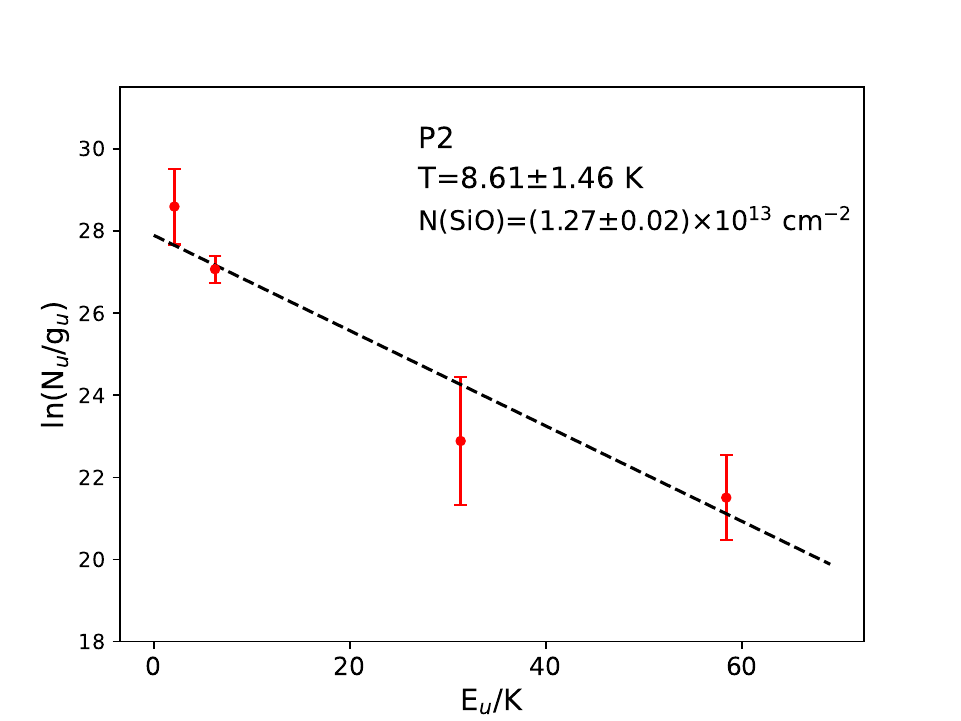}\quad \includegraphics[scale=0.36]{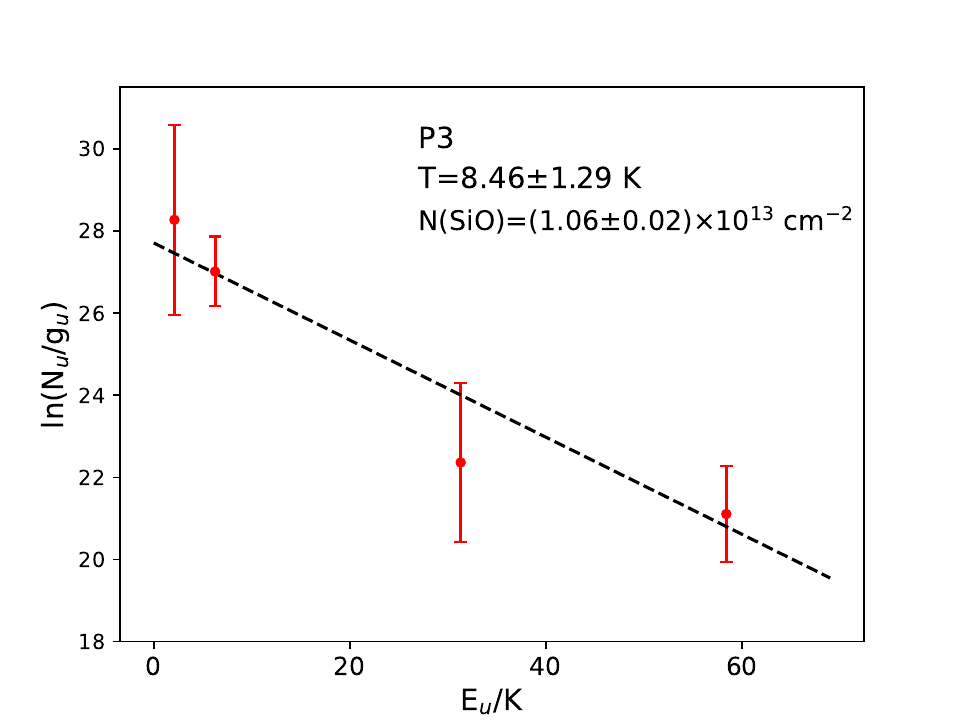}\quad\includegraphics[scale=0.36]{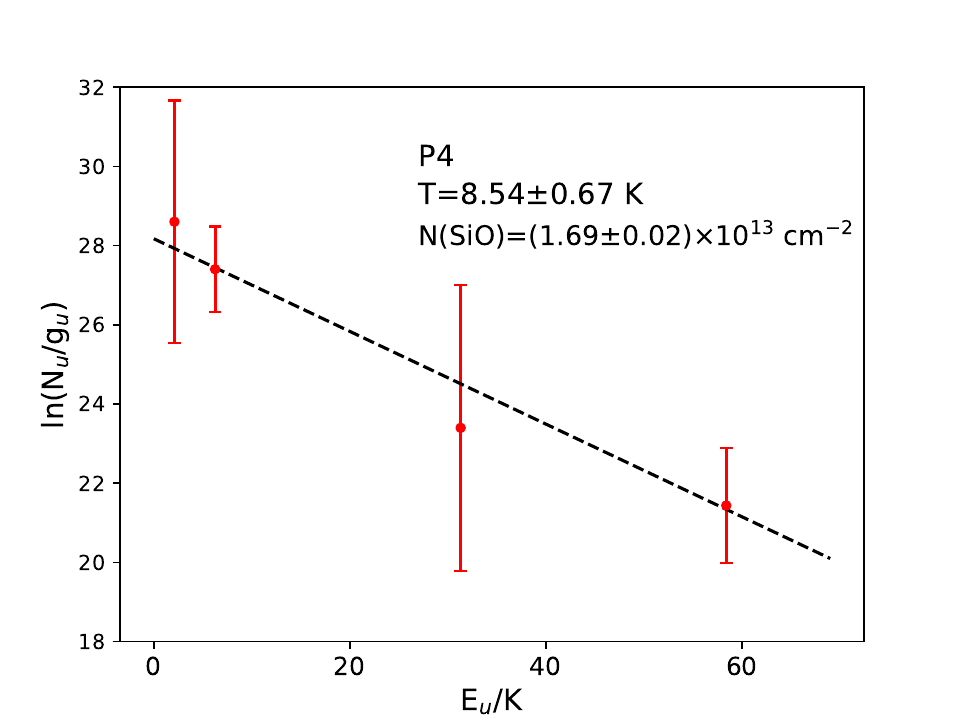}\quad\includegraphics[scale=0.36]{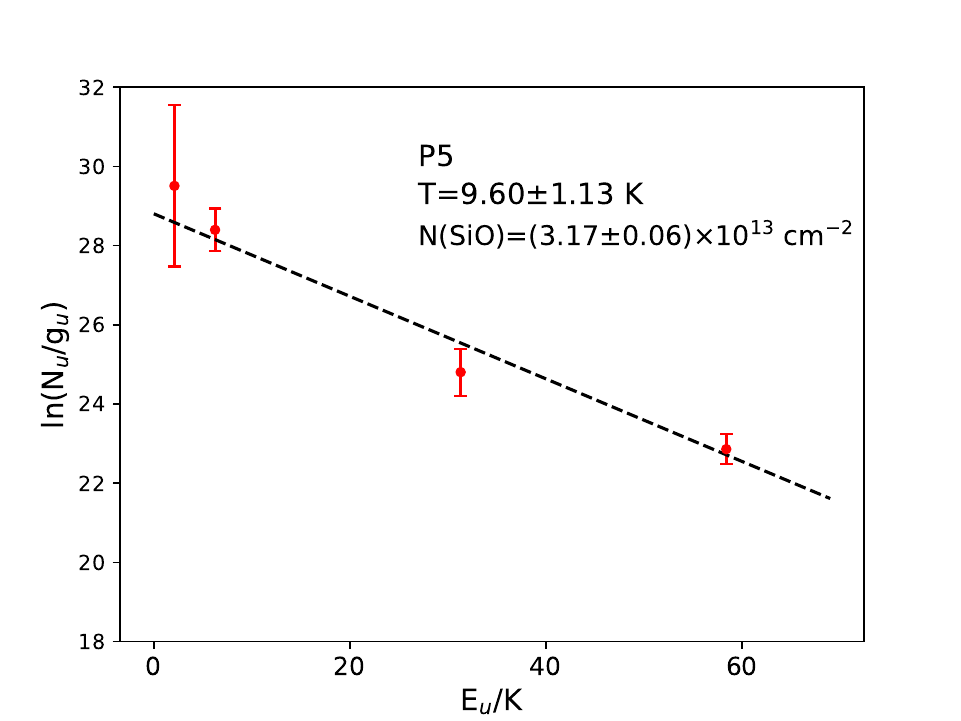}
\caption{Rotation diagrams for $J$= 1--0, 2--1, 5--4 and 7--6 transitions  toward positions P1, P2, P3, P4 and P5.}
\label{sio_rot}
\end{figure*}

As mentioned in Section 3.4, the distribution of Hi-GAL sources reveals a clustering of sources in the region. Of these, two are protostellar clumps. In addition, we also searched for thermal radio sources within the expanding shell. We find a source GPSR5 0.431+0.262 from the VLA survey for compact radio sources near the GC by \citet{2008ApJS..174..481L}, marked in Fig.~\ref{hgal_rad}. The object is detected at 1 GHz and 5 GHz bands (angular diameter of 1.9$\arcsec$). The radio spectral index measurements by \citet{2008ApJS..174..481L} show that it is thermal in nature with a spectral index $\alpha=0.2$. The angular size comparable to the image resolution of the data ($2.4''\times1.3''$) suggests that the source remains unresolved. At the GC distance, this gives an upper limit diameter of 0.1~pc. Assuming the object to be an \hii~region, we estimate the Lyman continuum photon rate of the \hii~region under the assumptions of optically thin emission and negligible absorption by the dust \citep{{1967ApJ...147..471M},{2016A&A...588A.143S}}
\begin{equation}
\left( {\frac{\rm{N}_{Lyc}}{\rm s^{-1}}}\right) = 4.771\times10^{42} \left( \frac{S_\nu}{\rm Jy}\right)\left( \frac{T_e}{\rm K}\right)^{-0.45} \left( \frac{\nu}{\rm GHz}\right)^{0.1} \left( \frac{d}{\rm pc}\right)^{2}
\label{nly}
\end{equation}
\noindent Here $S_{\nu}$ is the flux density at frequency $\nu$ taken as 26.7 mJy at 5 GHz \citep{1994ApJS...91..347B}, $T_e$ is the electron temperature,  and $d$ is the distance to the source that is 8.2 kpc. For estimating the electron temperature in the region, we use the electron temperature gradient curve across the Galactic disk \citep{2006ApJ...653.1226Q} and consider an approximate value of 7000~K at the GC. The resultant Lyman continuum photon rate is found to be $1.87\times10^{47}$~s$^{-1}$. Considering that the \hii~region is excited by a single zero age main sequence (ZAMS) star, we find the spectral type  \citep{1973AJ.....78..929P} of the source to be B0--B0.5. 

The source is also detected in the infrared bands (see Fig.~\ref{ircoshell}) as a compact source. However, no nebulous infrared emission is detected within the interior of the shell (see Fig.~\ref{ircoshell}). \hii~regions have characteristic infrared colors, that are useful for identifying these objects \citep[e.g.][]{{1989ApJ...340..265W},{1989AJ.....97..786H},{2012A&A...537A...1A}}. According to \citet{2017ApJ...846...64M}, the infrared colors log$_{10}(F_{24~\mu\textrm{m}}/F_{12~\mu\textrm{m}})\geq0$ and log$_{10}(F_{70~\mu\textrm{m}}/F_{12~\mu
\textrm{m}})\geq1.2$, and log$_{10}(F_{24~\mu\textrm{m}}/F_{12~\mu\textrm{m}})\geq0$ and log$_{10}(F_{160~\mu\textrm{m}}/F_{70~\mu\textrm{m}})\leq0.67$ reliably identify \hii~regions, independent of their size. To determine the infrared flux densities of the source, we use the MIPSGAL \citep{2015AJ....149...64G} and the Hi-GAL \citep{2016A&A...591A.149M} point source catalogues. For our source, we find log$_{10}(F_{24~\mu\textrm{m}}/F_{12~\mu\textrm{m}})$ = 0.9, log$_{10}(F_{70~\mu\textrm{m}}/F_{12~\mu\textrm{m}})=1.2$, and log$_{160}(F_{70~\mu\textrm{m}}/F_{12~\mu\textrm{m}})=0.4$, consistent with the \hii~region scenario. The identification of a potential \hii~region and the presence of clustered Hi-GAL clumps suggest that the shell's origin is likely due to feedback effects resulting from ongoing star formation activity. We will explore this scenario in detail in Section 4.1.

\subsection{SiO emission: evidence of large scale shocks}

To investigate the presence of shocks, in particular in view of the high Mach numbers discussed above, we carried out SiO observations toward 5 positions along the helix stream. SiO is a salient tracer of shocks as it is produced through the sputtering of Si-bearing material in grains \citep{1997A&A...321..293S}. The target positions are marked in Fig.~\ref{hgal_rad}. Four rotational transitions of SiO are observed and the details of these transitions are presented in Table~\ref{SiOtab}. The resultant SiO spectra are presented in Fig.~\ref{sio}. SiO emission is observed toward all five positions and all positions except P2 have two velocity components. Line parameters were extracted by fitting Gaussian components to the individual spectra and the results are tabulated in Table.~\ref{siofit}. A systematic velocity gradient is observed as we move from P1 to P5. For the $J=(2-1)$ transition, the line widths are in the range 10.3--30.9\kms with a mean value of 18.6\kms. Observed line widths are generally comparable to high velocity components observed in other Galactic clouds \citep[$\Delta \textrm V\geq$14\kms;][]{2016A&A...586A.149C}. To compare the properties of shocked gas emission with those of general dense gas, we examined  H$^{13}$CO$^+$ emission towards the same five positions (Fig.~\ref{sio}). The H$^{13}$CO$^+$ molecule serves as an excellent tracer for dense gas, given its critical density of $\sim10^5$ cm$^{-3}$ and the optically thin nature of its emission. The (2--1) transition of H$^{13}$CO$^+$ has a rest frequency of 86.75429~GHz, which is covered within the same spectral setup used for the SiO~(2--1) observations. Line parameters were extracted by fitting Gaussian components to the individual spectra and the results are tabulated in Table.~\ref{h13copfit}. Across all positions except P3, SiO line widths are generally broader than H$^{13}$CO$^+$ line widths. Specifically, the mean SiO (2-1) line width across all positions is 18.6\kms, while the mean H$^{13}$CO$^+$ (1-0) line width is 13.1\kms. The broader line width of SiO could be driven by moderate to high-velocity shocks and/or outflows. The exception at position P3 may reflect local variations in shock activity or gas dynamics.

\subsubsection{SiO column density and abundance}

In order to estimate the SiO column density and abundance in the helix stream, we follow the simplistic assumption that the emission is optically thin and assume local thermodynamic equilibrium (LTE) conditions. To compensate for difference in beam sizes and different beam dilution effects, we applied a correction factor $\theta_{1-0}^2/\theta_{line}^2$ to (2--1), (5--4) and (7--6) intensities, where $\theta_{1-0}$ is the beam FWHM of (1--0) transition and $\theta_{line}$ is the FWHM corresponding to the transition in consideration. Under these assumptions, the total SiO column density N(SiO) and the measured integrated line intensity $I$ in K\kms are related as \citep[e.g.,][]{2015MNRAS.446.3842A}

\begin{table}[!htbp]
\centering
\caption{Details of observed SiO transitions}
\begin{tabular}{c c c c c c}
\hline\\
Transition&Frequency &$E_\textrm{up}$&Log$_{10}$($A_\textrm{ij}$)&$S\,\mu^2$&rms\\
&(GHz)&(K)&(s$^{-1}$)&(D$^2$)&(mK)\\
\hline\\
1--0&43.4238&2.08 &	-5.51584&9.63&35\\
2--1&86.8469&6.25 &	-4.53354&19.26&20\\
5--4&217.1049&31.26&-3.28429&48.14&12\\
7--6&303.9268&58.35&-2.83461&67.38&4\\
\hline
\end{tabular}
\label{SiOtab}
\end{table}

\begin{equation}
    \frac{N_\mathrm{u}}{g_\mathrm{u}} = \frac{N(\mathrm{SiO})}{Q_\mathrm{rot}}e^{-E_\mathrm{u}/kT_\mathrm{rot}}=\frac{1.67\times10^{14}I}{\nu\mu^2S}
    \label{eqrot}
\end{equation}

\noindent where $N_\textrm{u}$ and $g_\textrm{u}$ are the molecular column density and degeneracy of the upper energy level, $Q_\textrm{rot}$ is the SiO rotational partition function, $E_\textrm{u}/k$ is the energy of the upper energy level 
(in K), $T_\textrm{rot}$ is the rotation temperature in K, $\nu$ is the transition frequency in GHz, $\mu$ is the dipole moment in Debye and $S$ is the line strength. We can construct a rotation diagram \citep[or ``Boltzmann plot'' ][]{GoldsmithLanger1999} by taking the natural logarithm of equation (\ref{eqrot}),

\begin{equation}
    \mathrm{ln}\frac{1.67\times10^{14}I}{\nu\mu^2S}=\mathrm{ln}\frac{\mathrm N(\mathrm{SiO})}{Q_{rot}}-\frac{E_u}{kT_{rot}}
\end{equation}

\noindent The SiO rotation diagrams for all five positions are presented in Fig.~\ref{sio_rot}. For positions exhibiting multiple velocity components, we summed all components to estimate rotation temperatures and column densities. We find rotation temperatures in the range 8.5--9.6~K and beam averaged column densities in the range $1.1-3.4\times10^{13}$~cm$^{-2}$ with a mean of $2.1\times10^{13}$~cm$^{-2}$. The column densities are listed in Table~\ref{SiOres}. We would like to caution that the temperature and column density estimates based on rotation diagram method could suffer from optical depth and beam dilution effects. It is worth noting that temperatures derived from rotation diagrams (8--10 K) are significantly lower than the typical kinetic temperatures expected in the CMZ \citep[25--200~K;][]{2017ApJ...850...77K}. This discrepancy is likely due to SiO lines becoming increasingly sub-thermal with increasing $J$ (and therefore increasing critical density), which is consistent with the presence of shocks.

\begin{figure}[t!]
\centering
\includegraphics[scale=0.54]{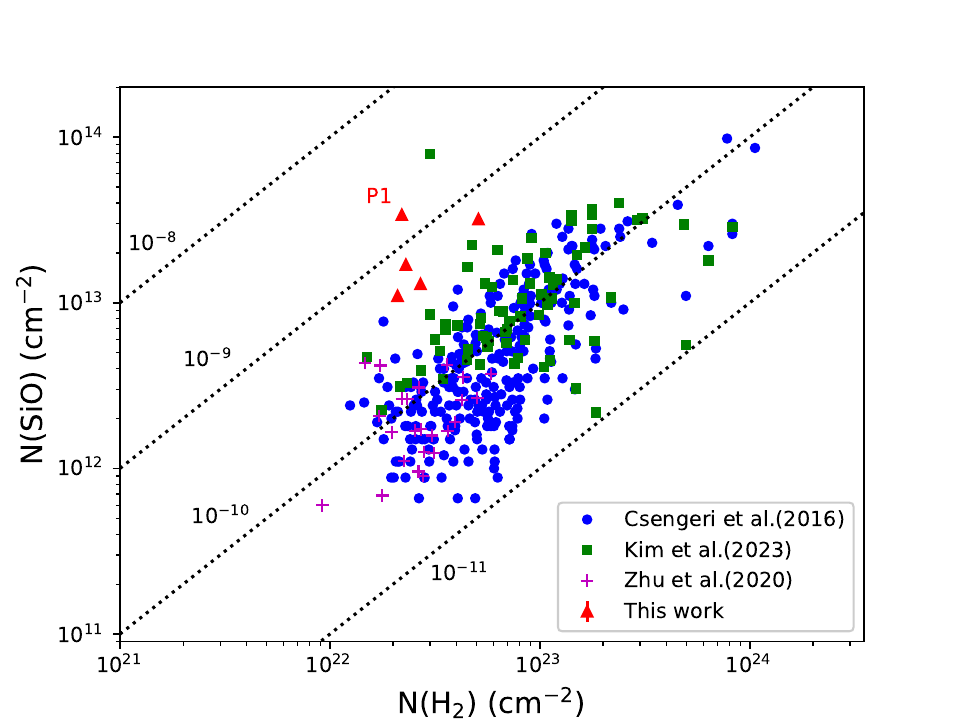}
\caption{SiO column density (N(SiO)) as a function of molecular hydrogen column density (N(H$_2$)). Dotted lines correspond to SiO abundances ($X$(SiO)) with respect to N(H$_2$). Red triangles correspond to helix stream, blue dots represent data from \citet{2016A&A...586A.149C}, magenta crosses from \citet{2020MNRAS.499.6018Z} and green squares correspond to data from \citet{2023A&A...679A.123K}.}
\label{sio_cden}
\end{figure}

\begin{figure}[t]
\centering
\includegraphics[scale=0.46]{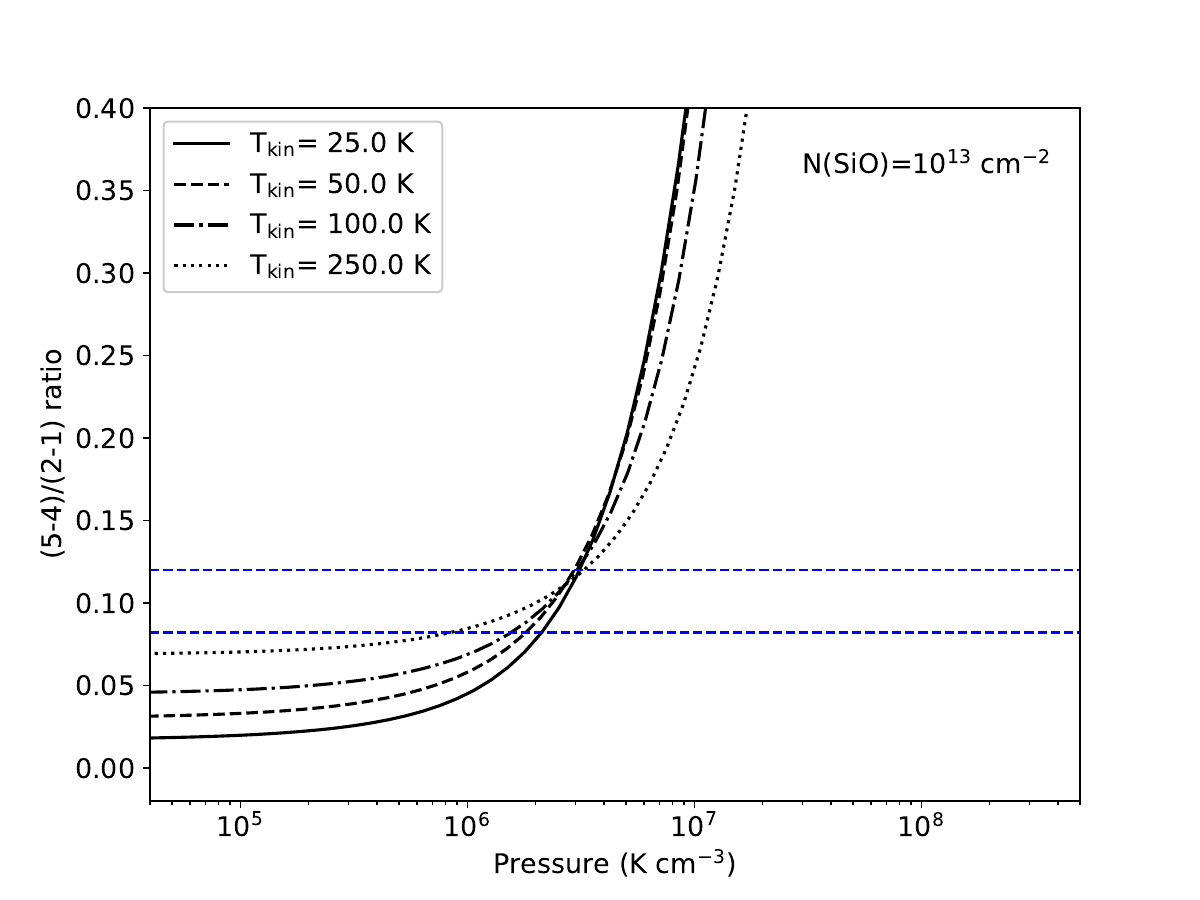}\quad\includegraphics[scale=0.46]{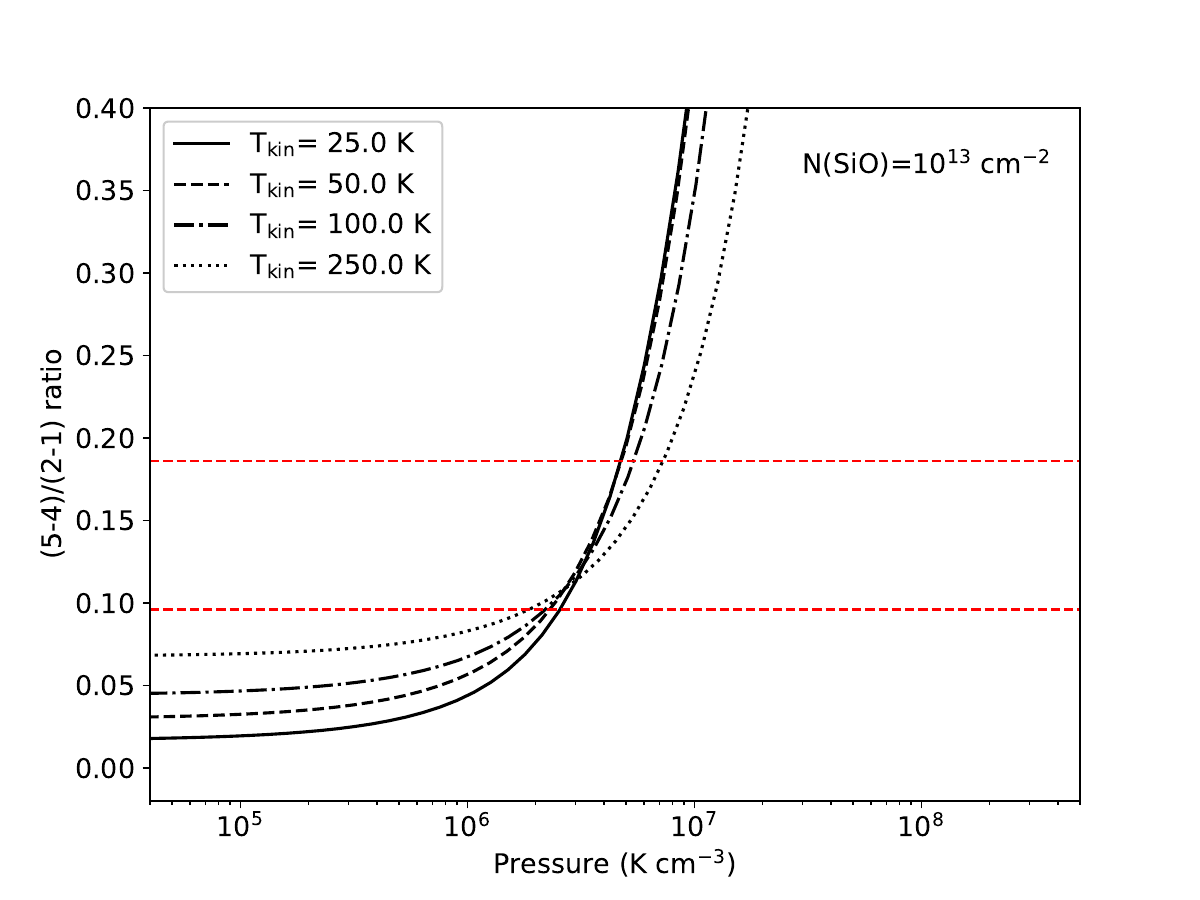}
\caption{RADEX predictions for the SiO~(5--4)/(2--1) integrated intensity ratio plotted as a function of the thermal pressure (n(H$_2$)T) for two different line widths 13\kms(top) and 30\kms (bottom). The calculations were carried out for SiO column density of 10$^{13}$~cm$^{-2}$ and kinetic temperatures of 25, 50, 100 and 250 K. The observed range of line ratios for the narrow components (represented by blue dashed lines) and broad components (represented by red dashed lines) are also plotted.}
\label{sio_radex}
\end{figure}
In order to estimate the abundance of SiO, $X$(SiO), relative to molecular hydrogen ($N$(SiO)/$N$(H$_2$)), we used the H$_2$ column density map produced by the PPMAP method. The column densities and abundances are listed in Table~\ref{SiOres}. The SiO abundances range from $4.8\times10^{-10}$ to $1.5\times10^{-9}$ with a mean of $7.7\times10^{-10}$. Interestingly, we see an enhanced abundance towards P1 ($1.5\times10^{-9}$) that is almost three times that of the mean abundance toward other positions ($5.9\times10^{-10}$). The integrated intensity ratios of SiO with respect to H$^{13}$CO$^+$ also shows a similar trend where the ratio towards P1 is thrice that of mean ratio towards other positions (see Table~\ref{SiOres}). P1 corresponds to the region where signatures of cloud-cloud collision are observed in the PV diagram and the PPV plot. Hence, the enhanced SiO abundance in this region could indicate SiO enrichment from cloud-cloud collision. In Fig.~\ref{sio_cden}, we compare the SiO and H$_2$ column densities of the helix stream with SiO data from other Galactic clouds. For this, we used the SiO column densities from \citet{2016A&A...586A.149C}, \citet{2020MNRAS.499.6018Z} and \citet{2023A&A...679A.123K}. We use the corresponding molecular hydrogen column density data from the ATLASGAL catalogue \citep{{2018MNRAS.473.1059U},{2022MNRAS.510.3389U}}. The sample consists of objects in different evolutionary stages such as infrared dark clouds, massive starless and star forming clumps, high mass protostellar objects, and ultracompact \hii~regions. From the figure, it is evident that the mean SiO abundance in the helix stream be higher than that of spiral arm clouds ($1.1\times10^{-10}$). However, the SiO abundance of the helix stream is similar to that of other GC clouds \citep[$\sim10^{-9}$, e.g.,][]{{1997ApJ...482L..45M},{2020MNRAS.499.4918A}}.

\subsubsection{Excitation conditions using radiative transfer modelling}

Given the limitations of the rotation diagram method in estimating the physical conditions of the helix stream, particularly due to the sub-thermal excitation of SiO lines and the presence of shocks, we employ the non-LTE modelling for a more reliable analysis. For this, we use the non-LTE radiative transfer code RADEX \citep{2007A&A...468..627V} that utilises the large velocity gradient (LVG) approximation. To determine SiO column densities, we compared the intensity ratios of the SiO~(2--1) and (5--4) lines, taking advantage of their comparable spatial resolutions ($\sim29''$). In Fig.~\ref{sio_radex}, we present RADEX non-LTE predictions illustrating the SiO~(5--4)/(2--1) intensity ratio as a function of thermal pressure (n(H$_2$)T) across a range of kinetic temperatures. We kept the SiO column density fixed at 10$^{13}$ cm$^{-2}$. Two separate plots are provided: one for $\Delta$V = 13 km/s, representing observed narrow components, and another for $\Delta$V = 30 km/s to account for broader components. From the plots, we find that the (5--4)/(2--1) line ratio as a function of thermal pressure remains nearly constant despite changes in kinetic temperature. The thermal pressures lie in the range 1.8--$7.3\times10^6$~K\,cm$^{-3}$. This pressure estimate from the RADEX analysis is consistent with the external pressure required for pressure-bounded equilibrium in the helix stream, evident from Fig.~\ref{radsig}(Bottom), and as discussed in Section 3.2.4. The relatively constant behaviour of the (5--4)/(2--1) line ratio with respect to thermal pressure suggests a decoupling of this ratio from changes in kinetic temperature which is indicative of non-LTE conditions in the helix stream. If we assume kinetic temperature of 50~K, typical for the CMZ, we get corresponding H$_2$ volume densities in the range 3.6--$14\times$10$^4$~cm$^{-3}$.

\begin{table}[!htbp]
\centering
\caption{SiO column densities and relative abundance}
\begin{tabular}{c c c c c}
\hline\\
&N(SiO)&N(H$_2$)&$X$(SiO)&$\displaystyle\frac{\textrm{I}_{\textrm{SiO~(2--1)}}}{\textrm{I}_{\textrm{H}^{13}\textrm{CO}^+(2-1)}}$\\
&cm$^{-2}$&cm$^{-2}$&&\\
\hline\\
P1&$3.4\times10^{13}$&$2.2\times10^{22}$&$1.5\times10^{-9}$&12.4\\
P2&$1.3\times10^{13}$&$2.7\times10^{22}$&$4.8\times10^{-10}$&4.6\\
P3&$1.1\times10^{13}$&$2.1\times10^{22}$&$5.2\times10^{-10}$&3.5\\
P4&$1.7\times10^{13}$&$2.3\times10^{22}$&$7.4\times10^{-10}$&4.2\\
P5&$3.2\times10^{13}$&$5.1\times10^{22}$&$6.3\times10^{-10}$&3.8\\
\hline

\end{tabular}
\label{SiOres}
\end{table}
\section{Discussion}

\subsection{Origin of the expanding CO shell}

Circular or shell-like cavities extending to parsec scales, termed as bubbles, are commonly observed within Galactic high mass star forming regions \citep{1977ApJ...218..377W} as well as in external galaxies \citep{{2023ApJ...944L..22B},{2023A&A...676A..67W}}. These bubbles are believed to be three-dimensional structures formed by radiation, spherical outflows or winds emanating from high mass stars \citep{2006ApJ...649..759C}. Such bubbles are often traced by infrared dust emission, atomic (HI) emission or molecular line emission \citep[e.g.,][]{{2003IAUS..212..596C},{2007ApJ...670..428C},{2010ApJ...709..791B},{2023ApJ...944L..22B}}. The velocity data and the morphological distribution of the molecular gas provide insights into the geometry of the shell and the thickness of the parent cloud. Shells can be classified into momentum driven or pressure driven based on the mechanism responsible for the expansion. Momentum driven shells are formed by stellar winds with sufficient strength where expansion is triggered by the transfer of wind mechanical energy into the ambient material \citep{1975ApJ...200L.107C}. In regions where weak or negligible stellar winds exist, the energy generated within an \hii~region could lead to a pressure difference between the ionised and the ambient neutral gas. This triggers a pressure-driven expansion of the \hii~region. As a consequence, a shock front forms at the boundary between the ionised gas and the ambient neutral gas and the shell traces the boundary of the shock \citep{1999PASP..111.1049G}.
\begin{figure}[t]
\centering
\includegraphics[scale=0.41]{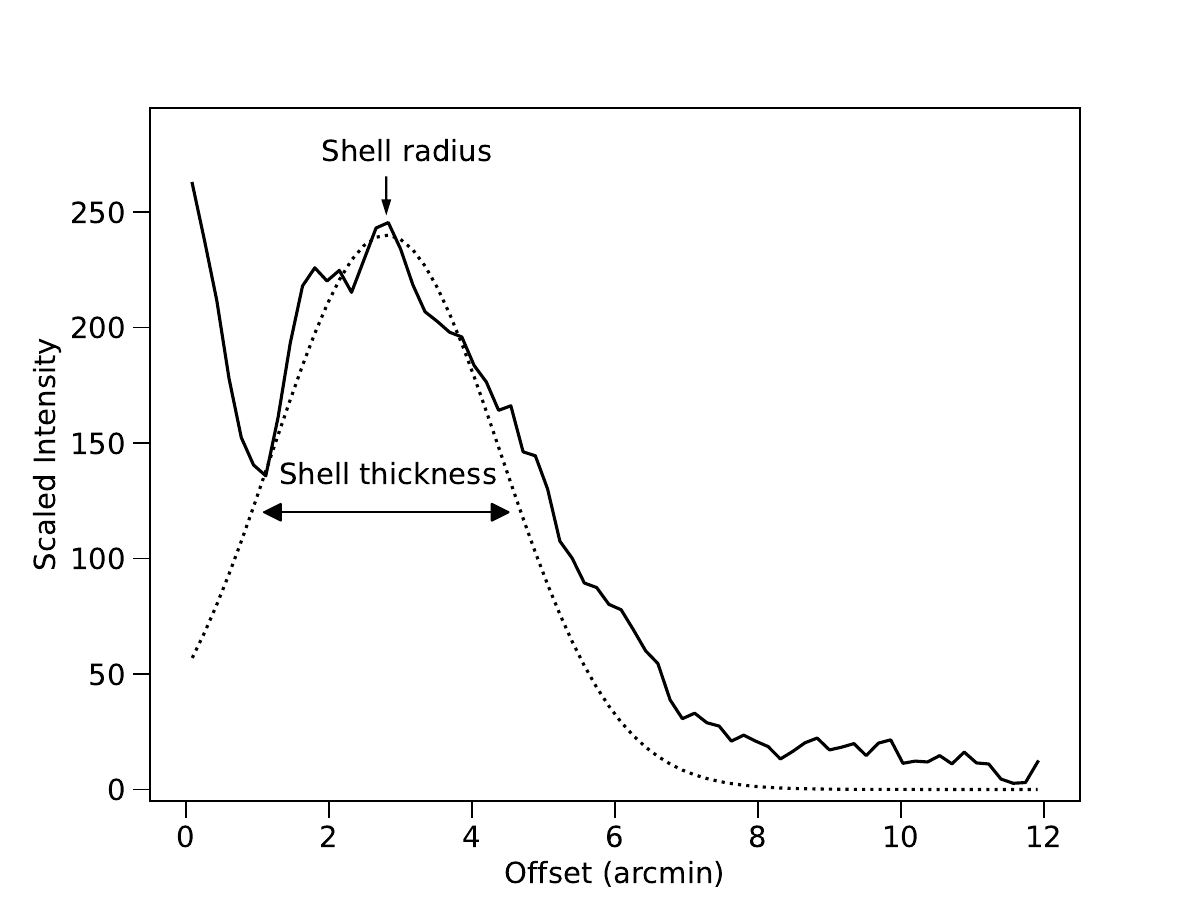}
\caption{Azimuthally averaged radial profile of the CO shell. Dotted line corresponds to the Gaussian fit to the intensity profile.}
\label{radprof}
\end{figure}
The expanding shell feature within the helix stream appears as a complete ring in the PV and PPV plots revealing blue-shifted and red-shifted emission corresponding to front and back hemispheres of the shell. This implies that the centre of the shell coincides with the centre of the cloud. The azimuthally averaged radial profile of the shell is presented in Fig.~\ref{radprof}. We estimated the radius and the thickness of the shell by fitting a Gaussian to the radial profile \citep[e.g.,][]{2011ApJ...742..105A}. The peak of the Gaussian corresponds to the radius of the shell whereas the FWHM corresponds to the thickness of the shell. We find the radius of the shell to be $2.8'$, equivalent to 6.7 pc, and the shell thickness to be $3.8'$, that is 8.9 pc. As there is no diffuse, large scale ionised or warm dust emission present within the shell (see Fig.~\ref{ircoshell}), we assume the source to be momentum driven, powered by stellar winds. Following the approach of \citet{2011ApJ...742..105A}, we test this hypothesis by estimating the wind mass loss rate ($\dot m_w$) required to generate the observed shell using the expression

\begin{equation}
\dot m_w = \frac{P_\mathrm{shell}}{v_w\,\tau_w}  
\end{equation}

\noindent where $P_\textrm{shell}$ is the total shell momentum, $v_w$ is the wind velocity and $\tau_w$ is the wind timescale. The momentum of the shell is given by $P_\textrm{shell}$ = $M_\textrm{shell}$\,$V_\textrm{exp}$ where $M_\textrm{shell}$ is the mass of the shell and $V_\textrm{exp}$ is the expansion velocity. Using the column density map based on PPMAP method, we find the mass of the shell as $1.4\times10^5$~M$_\odot$. Using the shell expansion velocity of 35\kms from the PV diagram (Fig.~\ref{pv}), we estimate $P_\textrm{shell}$ as $4.9\times10^6$~M$_\odot$\kms. Considering the radio source GPSR5 0.431+0.262 to be the source powering the stellar wind, we assume $v_w$=2000\kms for a typical O or B-type star \citep{{1998ASPC..131..218P},{2013ApJ...769L..16C}}. Assuming a wind timescale of 1 Myr, we find the mass loss rate as $2.5\times10^{-3}$~M$_\odot$yr$^{-1}$, notably exceeding the expected rate for O--B stars \citep[$10^{-7}-10^{-6}$~M$_\odot$yr$^{-1}$;][]{{2013ApJ...769L..16C}} by a factor of $\sim10^3-10^4$. This discrepancy strongly indicates that the observed shell formation is instigated by multiple sources or a highly energetic event. The momentum injected to the ISM by a single supernova explosion is $\sim10^5$~M$_\odot$\kms \citep{2020ApJ...905...35K}. This suggests that the shell might have been formed by multiple supernovae or a single hypernova, which releases $\sim10$ times more kinetic energy than a supernova \citep{1998Natur.395..672I}. This possibility aligns with similar observations in circumnuclear regions, such as the CO~0.13--0.13 cloud in the GC \citep{2001PASJ...53..779O} and the centre of the NGC 253 galaxy \citep{2006ApJ...636..685S}, where hypernova events have been proposed as potential candidates for driving molecular bubble/cavity expansions. We also detect seven X-ray sources from the $Chandra$ catalogue \citep{2006ApJS..165..173M} within the shell. These sources could potentially be cataclysmic variables, accreting neutron stars or black holes, young isolated pulsars, or Wolf-Rayet/O stars in colliding wind binaries. However, the specific identification of these sources and their association with the observed shell lie beyond the scope of the present study.

\subsection{Origin of SiO emission}

SiO emission could trace high velocity shocks ($v_\textrm{shock}\sim20-50$\kms) from protostellar outflows, medium to high velocity irradiated shocks ($v_\textrm{shock}\sim10-20$\kms to $v_\textrm{shock}\sim50$\kms) in hot cores, photon dominated regions and supernova remnants, and very low velocity shocks ($v_\textrm{shock}<10$\kms) in star forming regions and infrared dark clouds \citep{{2001A&A...372..281H},{2001A&A...372..291S},{2008A&A...490..695G},{2013ApJ...775...88N},{2016A&A...586A.149C},{2022MNRAS.511..953C},{2023A&A...679A.123K}}. Previous studies identified large scale SiO emission in the GC clouds associated with the SgrA region and the CMZ \citep[e.g.,][]{{1997ApJ...482L..45M},{2010A&A...523A..45R},{2012MNRAS.419.2961J}}. Typical SiO abundances are found to be $X_\textrm{SiO}\sim10^{-9}$. The enhanced SiO emission observed in the SgrA region exhibits a correlation with the 6.4 keV Fe line, as reported by \citet{2009ApJ...694..943A}. This correlation is attributed to two potential mechanisms: fluorescence in an X-ray reflection nebula or the impact of low-energy cosmic rays followed by electronic relaxation. SiO mapping studies toward the SgrB2 star forming region by \citet{2020MNRAS.499.4918A} reveal shocked gas with a turbulent substructure, consistent with a large-scale cloud-cloud collision. The average SiO abundance is $\sim10^{-9}$ and this along with observed SiO~(2--1) line intensities and gas temperature of $\sim30$~K agree with models of grain sputtering by C-type shocks.
\begin{figure}[!tbp]
\centering
\includegraphics[scale=0.25]{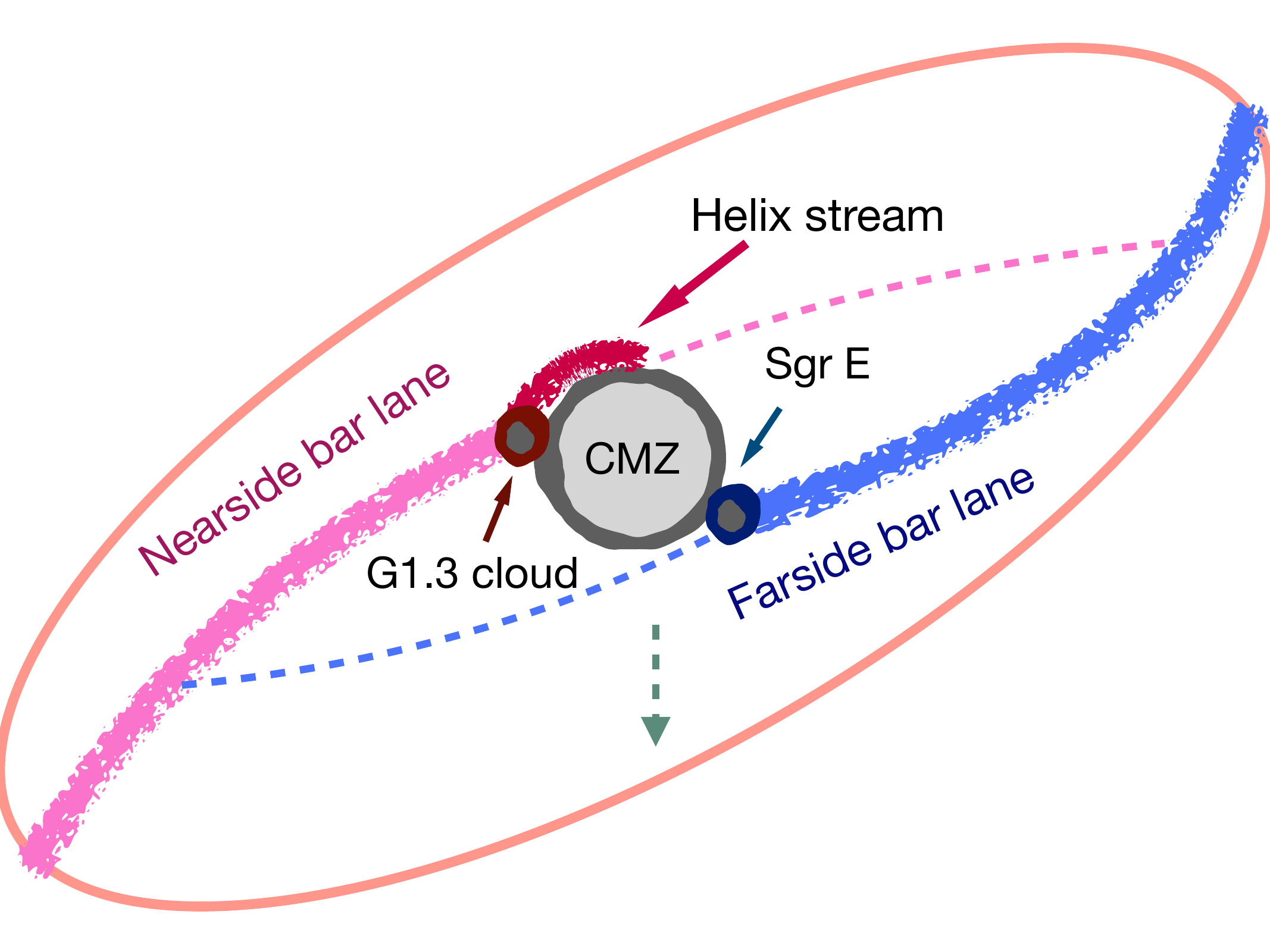}
\caption{Schematic view of the inner few kiloparsecs of the GC based on Fig.~3 of \citet{2023ASPC..534...83H}, showing near and far bar lanes, the CMZ, the G1.3 cloud complex, and the Sgr E complex. The relative position of the helix stream is also shown in the figure. Dashed blue and pink lines correspond to overshot gas from the near and far bar lanes, respectively. Dashed arrow indicate the direction of the observer's line of sight.}
\label{schematic}
\end{figure}

The consistent detection of SiO emission towards all five positions within the helix stream suggests the existence of a pervasive, large-scale shock distributed along the 200~pc structure. Among the many studies focused on SiO in the GC and the CMZ region, the helix stream holds a unique distinction. Its location $\sim50$~pc above the Galactic plane and the CMZ and its high LSR velocities sets it apart as an intriguing and singular feature within the GC. The width of the SiO line is related to the shock velocity and depending on the projection effects and turbulence, the shock velocity could be higher or lower than the line width \citep{2016A&A...595A.122L}. Considering the SiO~(2--1) line, we observe two components at all positions except P2. Specifically, positions P1, P4, and P5 display components with FWHM$>$20\kms. At P4 and P5, the FWHMs of the broader components range between 31--34\kms, three times larger than that of the narrowest component. Narrow components ($<$20\kms) identified toward all five positions have integrated intensities in the range 1.1--3.7~K\kms and is consistent with chemical models of low velocity shocks with n$_\textrm{H}=10^4$~cm$^{-3}$ \citep[$v_\textrm{shock}<20$\kms;][]{2016A&A...595A.122L}. Presence of broad components toward P1, P4 and P5 indicate the presence of moderate velocity shocks. This variation in shock velocities suggests diverse local conditions within the helix stream, indicating a complex environment with different shock strengths and possibly varied physical properties across different regions of the cloud.

The observed signatures of cloud-cloud collision towards the eastern edge of the cloud, coupled with indications of high supersonic turbulence, offer compelling evidence suggesting the formation of SiO associated with these events. These collisions are known to trigger shocks and compressive processes within the interstellar medium, creating conditions favourable to the formation of SiO. The presence of enhanced SiO abundance observed specifically at position P1 further supports the scenario of cloud-cloud collisions. This enhancement is consistent with the notion that the collision zone may have triggered sputtering of grains, elevating SiO levels. Regarding the broader components observed at positions P4 and P5, multiple factors could contribute to their origin. One plausible explanation could be the presence of heightened turbulence within these regions. Increased turbulence levels, possibly induced by dynamic processes such as collisions or shocks, could result in broader line widths. Alternatively, the broad components at P4 and P5 might also arise from localised star formation activities such as outflows, jets, and winds from young stellar objects as we observe multiple Hi-GAL clumps near P5. Follow up high resolution mapping studies, combined with simulations are crucial in unravelling the intricate interplay between cloud collisions, turbulence, outflows or jets of embedded sources in shaping observed SiO properties within the cloud.

\subsection{Formation of the helix stream}

In barred spiral galaxies, the central bar significantly shapes the dynamics of the gas reservoir within the galactic nucleus, crucially influencing their formation and evolution by funnelling matter from the spiral arms into the nuclear region \citep[e.g.,][]{{1985A&A...150..327C},{1990A&A...236..333H},{1993RPPh...56..173S},{2012ApJS..198....4O}}. Numerical simulations show that the gas that flows in and around the bar depends on periodic orbits and there exists two main classes of orbits: $x_1$ and $x_2$ orbits \citep{{1989A&ARv...1..261C},{1991MNRAS.252..210B},{1992MNRAS.259..345A}}. The $x_1$ orbits are elongated, with their major axis aligned along the bar direction whereas the $x_2$ orbits are innermost orbits, relatively less elongated and aligned perpendicular to the bar. The gas in outer regions follows the $x_1$ orbits, whereas the gas in the CMZ follows the $x_2$ orbits. The gas plunges from $x_1$ to $x_2$ orbits along the bar lanes. According to \citet{2023ASPC..534...83H}, the high velocity emission from the EMR/parallelogram originates from dense shocked overshooting gas that has been falling along the dust lane and is in the process of transitioning from $x_1$ to $x_2$ orbits. The shocked streams of gas also known as dust lanes are often observed in barred galaxies \citep{2023A&A...676A.113S} and fuel the star formation, accretion onto the supermassive black hole, and large-scale outflows \citep{2019MNRAS.484.1213S}. The time-averaged mass inflow rate along the dust lanes is $\sim2.7$~M$_\odot$yr$^{-1}$ out of which, up to 30$\%$ accretes onto the CMZ, while the rest overshoots and accretes later \citep{{2019MNRAS.484.1213S},{2021ApJ...922...79H}}. 

\begin{figure}[t]
\centering
\hspace*{-0.5cm}
\includegraphics[scale=0.24]{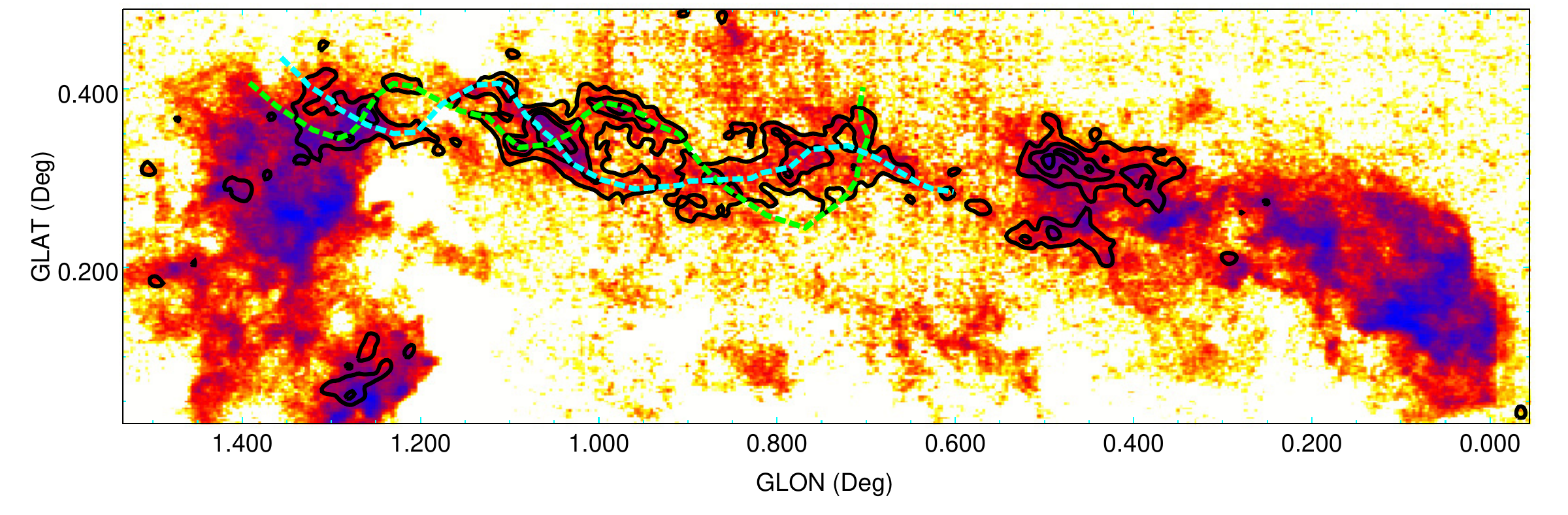}
\caption{Integrated intensity map of the $^{13}$CO emission in the velocity range 100, 200\kms (same as Fig.~\ref{13co}(Right)) overlaid with the $^{13}$CO integrated intensity contours in the velocity range 150, 200\kms, highlighting the helix morphology. Dashed green and cyan curves corresponds to two strands of the helix, strand 1 and strand 2.}
\label{helixs1s2}
\end{figure}
\begin{figure*}[t]
\centering
\includegraphics[scale=0.5]{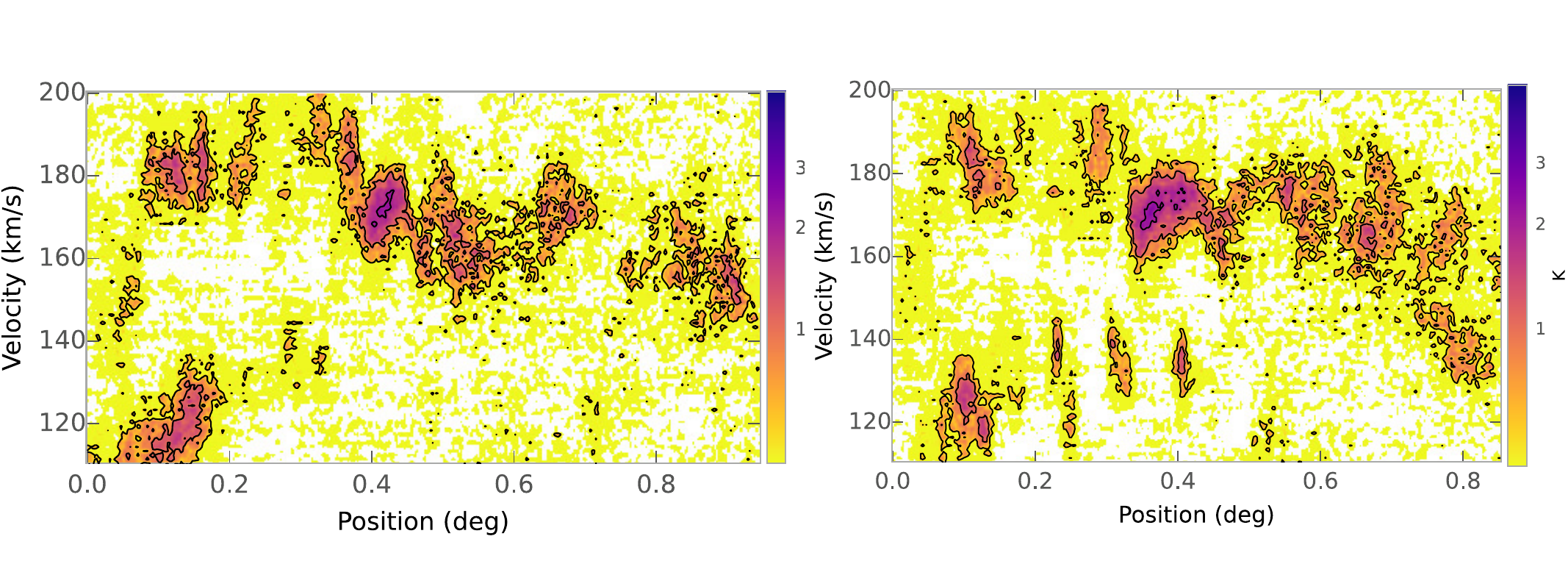}
\caption{(Left) PV diagram along the strand 1 of the helix stream shown as green curve in Fig.~\ref{helixs1s2}. (Right) PV diagram along the strand 2 of the helix stream shown as cyan curve in Fig.~\ref{helixs1s2}.}
\label{PVS1S2}
\end{figure*}
In the GC, there exists a population of compact clouds known as extended velocity features (EVFs) that are characterised by extreme velocity dispersions \citep[$\Delta \textrm{V}>100$\kms;][]{2019MNRAS.488.4663S}. According to \citet{{2019MNRAS.488.4663S}} and \citet{{2006A&A...447..533L},{2008A&A...486..467L}}, these are either (a) material which originates from collisions between the material in the dust lane and the material that has overshot from dust lanes on the opposite side \citep[e.g.,][]{2023ApJ...959...93G}, or (b) material that originates from the collision between the dust lane and the CMZ. \citet{2019MNRAS.488.4663S} identified an EVF at $l=1.3^\circ$, which is proposed to be of the second category, i.e., collision between the dust lane and the CMZ. 

The G1.3 cloud ($l\sim1.28^\circ$, $b\sim0.07^\circ$) is located at the edge of the CMZ and has high SiO abundance \citep[$2.6\times10^{-9}$;][]{2018A&A...613A..42R}. \citet{2022A&A...668A.183B} investigated the morphology, physical properties as well as the chemical composition of the G1.3 cloud. They found signatures of cloud-cloud collisions, consistent with gas accretion from the near-side dust lane onto the CMZ region. Two prominent velocity components ($\sim$100, 180\kms) are observed, connected by an emission bridge. The eastern edge of the helix stream is at $l=1.3^\circ$, where we also observe signatures of cloud-cloud collisions in the PV and the PPV plots. It is located 40~pc above the G1.3 cloud. The similarity in velocity components ($\sim$127, 177\kms) and estimated SiO abundance towards position P1 at the eastern edge ($1.9\times10^{-9}$) suggests that the helix stream is physically connected to the G1.3 cloud. The helix stream likely originates from turbulent shocked gas resulting from the interaction between the near dust lane and the CMZ, overshot to higher latitudes (see Fig.~\ref{schematic}), while the other part collided with the CMZ, forming the G1.3 cloud. The helix stream extends beyond the Galactic plane, with the eastern edge indicating the bar lane brushing against the CMZ, forming a velocity bridge, and the rest of the stream continuing along the bar lane trajectory. On the opposite side, the Sgr E region is believed to be at the intersection of the far dust lane and the CMZ. \citet{2022ApJ...939...58W} identified CO filaments at much smaller scales ($\sim$2~pc) in the Sgr E region with aspect ratio $\sim5:1$, and explained them as gas being stretched as it is rapidly accreted by  the gravitational field of the Galactic bar while falling toward the CMZ.

The origin of the double helix morphology of the cloud is unclear. In order to examine the velocity structure of the helix stream in detail, we generated additional PV diagrams along two strands of the helix as depicted in Fig.~\ref{helixs1s2}. The resultant PV diagrams along strand 1 and strand 2 are illustrated in Fig.~\ref{PVS1S2}. Across both diagrams, we identify two cloud components. The primary component, indicative of the velocity of the helix stream, exhibits velocities exceeding 150\kms, while the secondary component, manifesting as clumpy features, aligns with the bridge feature (Fig.~\ref{pv}), associated with the cloud-cloud collision with the CMZ, with velocities lower than 140\kms. The PV diagrams unveil a distinct helical or cork-screw morphology at scales of 6~pc and 14~pc for strand 1 and strand 2, respectively, implying twisting and turning motions within the two strands of the helix stream. Similar twisting motions have been observed in few Galactic filamentary clouds \citep[e.g.,][]{{2019MNRAS.489.4771G},{2021ApJ...908...86A}}. Beyond these smaller scale helical features, the PV diagram towards strand 1 also exhibits a large-scale wave-like morphology with a characteristic wavelength of approximately $\sim$86~pc, signifying the complex kinematics of the helix stream. 

In addition to the helix stream, few other clouds within the GC region exhibit a helix or double-helix morphology. The double helix nebula (DHN) is a 25 pc infrared nebula that is $\sim$100~pc above the GC and is interpreted as a torsional Alfv\'en wave propagating vertically away from the Galactic disc, possibly driven by the rotation of the magnetised circumnuclear disc \citep{2006Natur.440..308M}. The GC molecular tornado (GCT) is a helical-spur of molecular gas observed in CO at $l=1.2^\circ$, V$_\textrm{LSR}$=70\kms, extending 170~pc vertically in the plane and is believed to be formed by magnetic squeezing mechanism where a vertical magnetic tube/flux squeezed by the molecular gas in Galactic rotation \citep{2007PASJ...59..189S}. The pigtail cloud is another helical CO cloud extending $\sim$30~pc vertically at $l=-0.7^\circ$, V$_\textrm{LSR}\sim$--70 to --30\kms and is proposed to be associated with a twisted and coiled magnetic tube resulting from interaction of clouds in the $x_1$ and $x_2$
orbits \citep{2012ApJ...756...87M}. Unlike these clouds, the helix stream is aligned nearly parallel to the Galactic plane. Interestingly, the GCT cloud is spatially located close to the eastern edge of the helix stream. From Fig.~\ref{pv}(Left), we observe bridging emission features from 70\kms, the velocity of GCT to 180\kms, the velocity of the helix stream. Its spatial proximity to the GCT, coupled with bridging emission features observed between their velocities, strongly suggests a physical connection between the two. This connection implies a significant linkage between the helix stream and the magnetically-induced GCT, potentially indicating a common mechanism shaping both clouds. A comparison of the $^{13}$CO emission from the helix stream with the dust polarisation measured with the Planck 353 GHz data \citep{2020A&A...641A..12P} is presented in Fig.~\ref{planckpol}. The observed dust polarisation, which is directly linked to the magnetic field of the cloud, reveals that the polarisation vectors are parallel to the high-intensity ridges within the helix stream. Consequently, the orientation of the magnetic field (rotated 90$^\circ$ with respect to the dust polarisation angle) is perpendicular to the high-intensity emission. However, the observed dust polarisation could potentially be contaminated by line-of-sight dust emission. Therefore, detailed follow-up polarisation studies are necessary to accurately disentangle the true magnetic field structure and its relation to the observed morphology of the helix stream.

An alternate proposition for the origin of the double helix morphology is that the two strands of the double helix represent two interacting high velocity gas streams that are gravitationally wound to each other, while moving along their respective trajectories and revolving around each other. Under this assumption, we can roughly estimate the mass of the streams by applying the Kepler's third law,

\begin{equation}
\frac{P^2}{a^3}=\frac{4\pi^2}{G(M_1+M_2)}   
\end{equation}

\noindent where $P$ is the orbital period, $a$ is the separation between the two streams of gas and $M_1$ and $M_2$ correspond to the masses of the individual streams and $G$ is the gravitational constant. We calculate $P$ by dividing the 2D wavelength of the helix ($\sim$30~pc) with the velocity of the streams. The line of sight velocity of the streams is $\sim$170\kms and assuming that the total velocity vectors point in the same direction as the bar major axis ($\sim$30$^\circ$ with respect to the line of sight), we find the stream velocity as 80\kms. Using these, we estimate P to be 0.36~Myr. Assuming $M_1$=$M_2$, and $a$=4.5~pc (half of the separation between the two strand of the helix), we find the total mass of the system as $5.9\times10^6$~M$_\odot$. We find that the calculated mass using Kepler's third law is comparable to the observed mass estimate for the helix stream ($\sim2-5\times10^6$~M$_\odot$). Large-scale oscillatory gas flows with wavelengths ranging from 0.3--400~pc are observed toward molecular clouds in Galactic and extragalactic environments. \citet{2020NatAs...4.1064H} proposed that they are likely to be formed via gravitational instabilities. \citet{2020Natur.578..237A} identified a coherent 2.7 kpc sinusoidal wavy chain of gas clouds in the Solar neighbourhood, known as the Radcliffe Wave. Follow-up studies by \citet{2024Natur.628...62K} show that the Radcliffe Wave is a coherent oscillating structure, with phase velocities ranging from 5--40\kms. Gravitational perturbations and feedback mechanisms have been suggested as the possible origins of the Radcliffe wave. Considering that the helix stream could also have its morphology and velocity structure shaped by gravitational instabilities, it is possible that the strands of the double helix can be approximated as two superposed Radcliffe Wave-like structures.

The intricate morphology and kinematics of the helix stream, along with its association with other helical clouds and the application of Kepler's third law, provide compelling evidence for a multifaceted origin involving gravitational interactions, magnetic fields, and possibly other physical processes. Our study revealing active massive star formation within this cloud for the first time not only challenges existing paradigms but also emphasise the significance of clouds with high line of sight velocities in the evolution of the Galaxy. Furthermore, the helix stream stands as an invaluable local template, offering insights into understanding sub-parsec scale mass inflow dynamics within the nuclear regions of barred galaxies.


\section{Conclusion}

We carried out a comprehensive multiwavelength study of the high velocity helix stream in the GC region, that is associated with the EMR/parallelogram. Using CO data,  we performed a kinematic analysis of the cloud, followed by an investigation of the star formation activity using data from infrared to radio wavelengths. We explored the presence of shocked gas in the region using SiO molecular line data. The outcomes of this study can be summarised as follows: \\

\begin{itemize}
    
\item{The helix stream is a highly elongated (aspect ratio: 9.3) 200 pc high velocity cloud (100--200\kms) positioned at a vertical distance of $\sim$15--55 pc above the CMZ. The mass of the cloud is estimated to be $2.5\times10^6$~M$_\odot$. }\\

\item{The $^{12}$CO PV and $^{13}$CO PPV plots of the cloud display a bridging feature towards the eastern edge connecting the helix stream to several clumpy features in the velocity range 70--100\kms, a signature of cloud-cloud collision. } \\

\item{The $\sigma-R$ plot of the helix stream is fitted with a power-law index of 0.7, that is steeper than the classical power-law index of 0.5. The obtained index is similar to that of other GC clouds. We find the gas motion within the cloud to be supersonic ($\mathcal{M}_{3D}>20$).} \\

\item{The PV and PPV plots also reveal a ring-like feature at $l\sim0.42^\circ$, associated with an expanding shell of molecular gas. We find the expansion velocity to be 35\kms. The shell radius and thickness are 6.7~pc and 8.9~pc, respectively. The large momentum of the expanding shell suggests its formation associated with multiple supernovae or a single hypernova.} \\

\item{A thermal radio source is identified towards the centre of the expanding CO shell. The infrared colours of the radio source is consistent with that of an \hii~region, revealing the massive star formation activity within the shell. The ZAMS spectral type of the ionising source is estimated to be B0--B0.5.}\\

\item{We find 19 Hi-GAL dust clumps within the cloud indicating the ongoing star formation activity within the cloud.  There are 8 prestellar clumps and 11 protostellar clumps. Of these clumps, 17 satisfy criteria for massive star formation.} \\

\item{Detection of SiO emission within the helix stream indicate the presence of a large-scale pervasive shock in the region.  Mean SiO abundance is $\sim10^{-9}$. An enhancement in SiO abundance is observed towards the eastern edge,  suggesting additional SiO production from a cloud-cloud collision.} \\

\item{The PV diagrams along the individual strands of the helix stream reveal twisting and turning motions within the cloud at scales of 6--14 pc. In addition, the PV diagrams also reveal larger scale ($\sim$86~pc) wavy patterns.} \\

\item{The helix stream is likely formed by the dust lane-CMZ interaction and represents the turbulent shocked gas that is overshot to high latitudes after ``brushing'' the CMZ at the location of G1.3 cloud.  This interpretation finds support in the observed kinematic signatures, as well as large Mach numbers and enhanced SiO abundance within the cloud. The distinctive double helix structure of the cloud may arise from the combined influences of turbulence, gravity, and magnetic squeezing.}\\

\end{itemize}

\begin{acknowledgements}

We thank the referee for the valuable comments and suggestions that improved the quality of the paper. We thank the staff of APEX 12m telescope, IRAM 30m telescope and Yebes 40m telescope for their excellent support. This publication is based on data acquired with the Atacama Pathfinder Experiment (APEX) under programmes 092.F--9315, 193.C--0584, 109.A--9518, and 109.C--9518. APEX is a collaboration among the Max-Planck-Institut fur Radioastronomie, the European Southern Observatory, and the Onsala Space Observatory. The processed data products are available from the SEDIGISM survey database located at https://sedigism.mpifr-bonn.mpg.de/index.html, which was constructed by James Urquhart and hosted by the Max Planck Institute for Radio Astronomy. This work is based on observations carried out with the Yebes 40 m telescope (21A017). The 40 m radio telescope at Yebes Observatory is operated by the Spanish Geographic Institute (IGN; Ministerio de Transportes y Movilidad Sostenible). This work is based on observations carried out under project number 057--21 with the IRAM 30m telescope. IRAM is supported by INSU/CNRS (France), MPG (Germany) and IGN (Spain). This work is based in part on observations made with the Spitzer Space Telescope, which was operated by the Jet Propulsion Laboratory, California Institute of Technology under a contract with NASA. This publication also made use of data products from Herschel. Herschel is an ESA space observatory with science instruments provided by European-led Principal Investigator consortia and with important participation from NASA. This
work has made use of 353 GHz dust polarisation data, based on
observations obtained with Planck (http://www.esa.int/Planck), an
ESA science mission with instruments and contributions directly
funded by ESA Member States, NASA, and Canada. MCS acknowledges financial support from the European Research Council under the ERC Starting Grant ``GalFlow'' (grant 101116226). D. Riquelme acknowledges the financial support of DIDULS/ULS, through the project PAAI 2023.

\end{acknowledgements}

\bibliography{ref}
\bibliographystyle{aa}
\clearpage
\newpage
\onecolumn
\appendix
\section{Additional figures and tables}

\begin{sidewaystable}[!htbp]
\centering
\caption{SiO line parameters}
    \begin{tabular}{c c c c c c c c c c c c c}
 \hline
    \hline\\
&\multicolumn{3}{c}{SiO (1--0)}&\multicolumn{3}{c}{SiO~(2--1)}&\multicolumn{3}{c}{SiO~(5--4)}&\multicolumn{3}{c}{SiO (7--6)}\\
    \hline\\
    
&$\int \textrm T_{MB}\,\textrm{dv}$& V$_\textrm{LSR}$& $\Delta \textrm V$&$\int \textrm T_{MB}\,\textrm{dv}$& V$_\textrm{LSR}$& $\Delta \textrm V$&$\int \textrm T_{MB}\,\textrm{dv}$& V$_\textrm{LSR}$& $\Delta \textrm V$ &$\int \textrm T_{MB}\,\textrm{dv}$& V$_\textrm{LSR}$& $\Delta \textrm V$\\

&(K\,km~s$^{-1}$)&(km~s$^{-1}$)& (km~s$^{-1}$) &(K\,km~s$^{-1}$)&(km~s$^{-1}$)& (km~s$^{-1}$)&(K\,km~s$^{-1}$)&(km~s$^{-1}$)& (km~s$^{-1}$) &(K\,km~s$^{-1}$)&(km~s$^{-1}$)& (km~s$^{-1}$)\\
\hline\\

P1&$9.6\pm0.3$&$126.4\pm0.3$&$23.9\pm0.8$&$5.5\pm0.1$&$126.5\pm0.2$&$23.6\pm0.5$ &$0.5\pm0.06$&$127.8\pm1.4$&$22.2\pm2.6$&$0.14\pm0.01$&$129.8\pm0.8$&$14.3\pm1.7$\\
&$5.8\pm0.2$&$177.3\pm0.3$&$16.1\pm0.7$&$3.7\pm0.1$&$177.0\pm0.2$&$18.3\pm0.5$&$0.4\pm0.05$&$176.8\pm1.1$&$16.1\pm2.4$&$0.01\pm0.02$&$181.3\pm0.8$&$12.9\pm2.1$\\
\hline\\

P2&$6.6\pm0.2$&$168.5\pm0.2$&$15.4\pm0.6$&$3.1\pm0.1$&$168.4\pm0.2$&$16.5\pm0.5$&$0.3\pm0.03$&$168.3\pm0.7$&$11.1\pm2.0$ &$0.07\pm0.01$&$167.4\pm1.1$&$9.9\pm2.3$\\
\hline\\

P3 &$4.8\pm0.4$&$156.2\pm1.6$&$37.6\pm3.9$&$1.8\pm0.1$&$149.2\pm0.3$&$10.3\pm0.8$&$0.2\pm0.02$&$150.2\pm0.5$&$6.2\pm1.4$&$0.04\pm0.01$&$146.2\pm1.3$&$11.0\pm3.3$\\
&-&-&-&$1.1\pm0.1$&$163.3\pm0.7$&$11.8\pm1.4$&-&-&-&-&-&-\\
\hline\\

P4&$4.8\pm0.5$&$133.7\pm0.9$&$19.9\pm2.6$&$1.5\pm0.2$&$132.1\pm0.3$&$10.5\pm1.0$&$0.2\pm0.02$&$131.3\pm0.9$&$7.1\pm2.1$&$0.06\pm0.02$&$138.0\pm1.9$&$14.9\pm6.2$\\
&$1.8\pm0.5$&$163.9\pm2.9$&$21.9\pm7.1$  &$2.8\pm0.3$&$145.7\pm1.4$&$30.9\pm2.0$&$0.3\pm0.1$&$140.2\pm3.3$&$15.8\pm4.9$&-&-&-\\
\hline\\

P5&$5.2\pm0.7$&$94.9\pm0.3$&$12.9\pm0.9$&$3.1\pm0.4$&$95.2\pm0.3$&$11.5\pm0.8$&
$0.3\pm0.04$&$96.1\pm0.4$&$6.8\pm0.9$&$0.03\pm0.01$&$98.8\pm0.5$&$3.9\pm1.0$\\
&$11.1\pm0.9$&$114.4\pm1.2$&$31.5\pm2.3$&$8.5\pm0.5$&$116.0\pm1.0$&$34.2\pm2.3$&$1.6\pm0.08$&$112.8\pm0.8$&$31.8\pm1.7$&$0.22\pm0.01$&$107.5\pm1.9$&$28.5\pm1.9$ \\
\hline 
\hline\\
\end{tabular}
\label{siofit}
\end{sidewaystable}

\begin{table}[!htbp]
\centering
\caption{Parameters of H$^{13}$CO$^+$(2--1) lines}
\begin{tabular}{cccc}
\hline
\hline
& $\int \textrm{T}_{\textrm{MB}}\,\textrm{dv}$ & V$_{\textrm{LSR}}$ & $\Delta \textrm{V}$ \\
& (K km s$^{-1}$) & (km s$^{-1}$) & (km s$^{-1}$) \\
\hline \\
P1 & $0.8 \pm 0.1$ & $121.8 \pm 1.3$ & $21.7 \pm 3.3$ \\
   & $0.07 \pm 0.06$ & $138.9 \pm 2.3$ & $5.8 \pm 3.5$ \\
   & $0.2 \pm 0.05$ & $179.6 \pm 1.1$ & $7.9 \pm 2.8$ \\
\hline \\
P2 & $0.7 \pm 0.1$ & $167.6 \pm 0.7$ & $12.8 \pm 1.7$ \\
\hline \\
P3 & $0.6 \pm 0.1$ & $150.7 \pm 0.9$ & $15.6 \pm 2.4$ \\
   & $0.2 \pm 0.1$ & $169.7 \pm 1.5$ & $9.9 \pm 3.2$ \\
\hline \\
P4 & $0.6 \pm 0.1$ & $131.1 \pm 4.1$ & $13.4 \pm 4.1$ \\
   & $0.5 \pm 0.1$ & $149.4 \pm 4.1$ & $16.1 \pm 4.1$ \\
\hline \\
P5 & $2.0 \pm 0.4$ & $95.4 \pm 1.3$ & $13.8 \pm 2.4$ \\
   & $1.1 \pm 0.4$ & $112.5 \pm 2.4$ & $13.9 \pm 5.4$ \\
\hline 
\hline
\end{tabular}
\label{h13copfit}
\end{table}

\begin{figure*}[h!]
\centering
\includegraphics[scale=0.7]{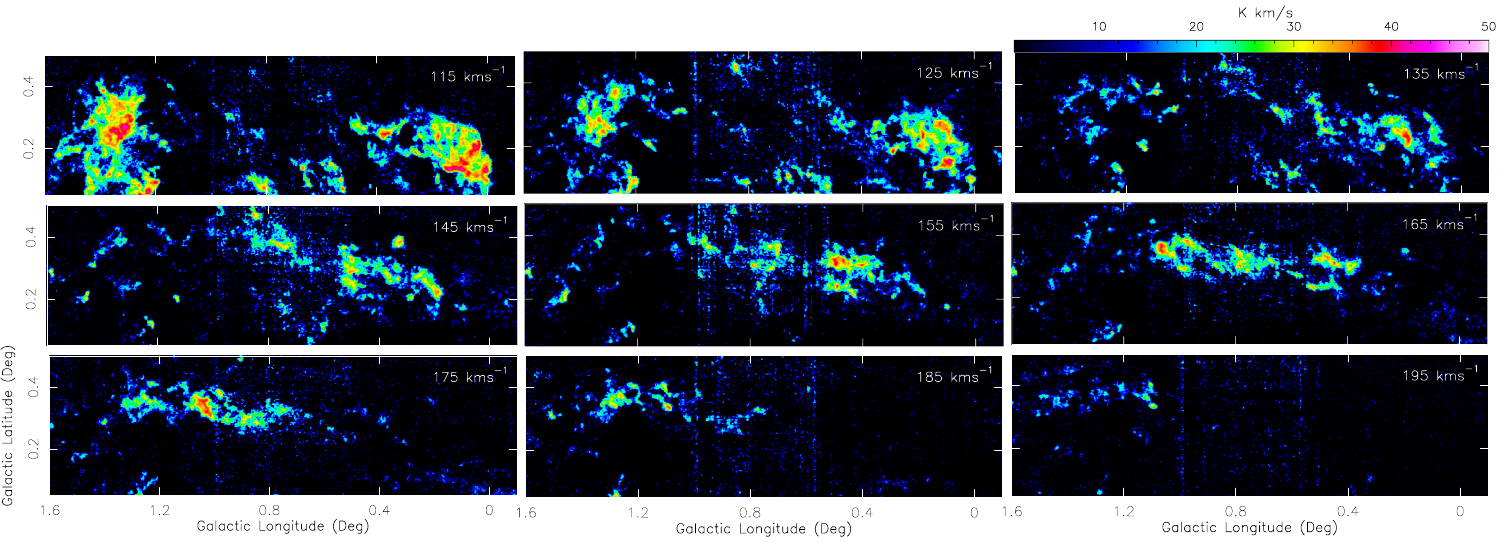}    
\caption{SEDIGISM $^{13}$CO intensity maps in 10 \kms velocity intervals from +110 to +200 km/s. The velocity labels in individual panels correspond to central velocities of each interval.}
\label{channel}
\end{figure*}
\begin{figure*}[h!]
\centering
\includegraphics[scale=0.33]{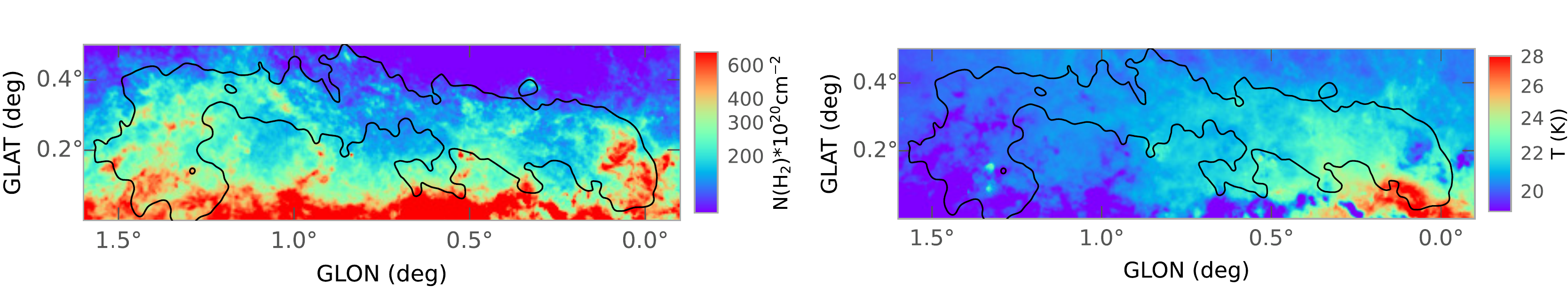}
\caption{(Left) Column density and (Right) dust temperature maps using PPMAP method overplotted with 5$\sigma$ contours of $^{13}$CO integrated emission.}
\label{Tcd}
\end{figure*}

\begin{figure*}[h!]
\centering
\includegraphics[scale=0.42]{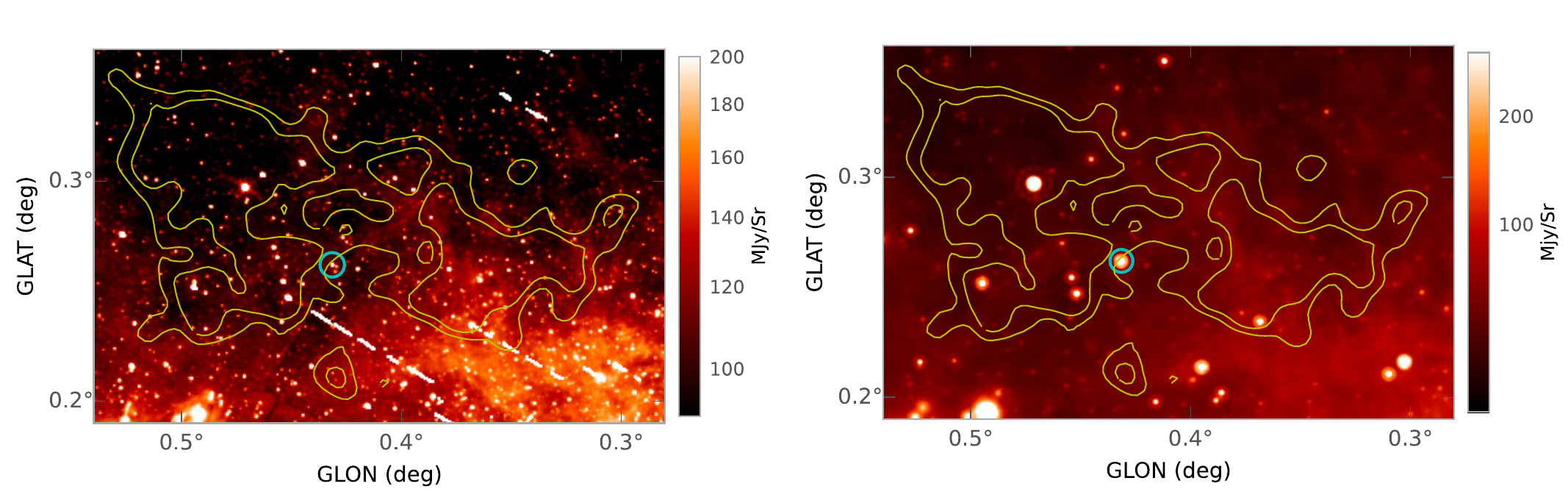}
\caption{Mid-infrared 8~$\mu$m (Left) and 24~$\mu$m (Right) maps of the expanding CO shell overlaid with $^{13}$CO emission contours. Contour levels are 18 and 23 Kkm/s. The thermal radio source GPSR5~0.431+0.262 is marked as a circle.}
\label{ircoshell}
\end{figure*}

\begin{figure*}[h!]
\centering
\includegraphics[scale=0.5]{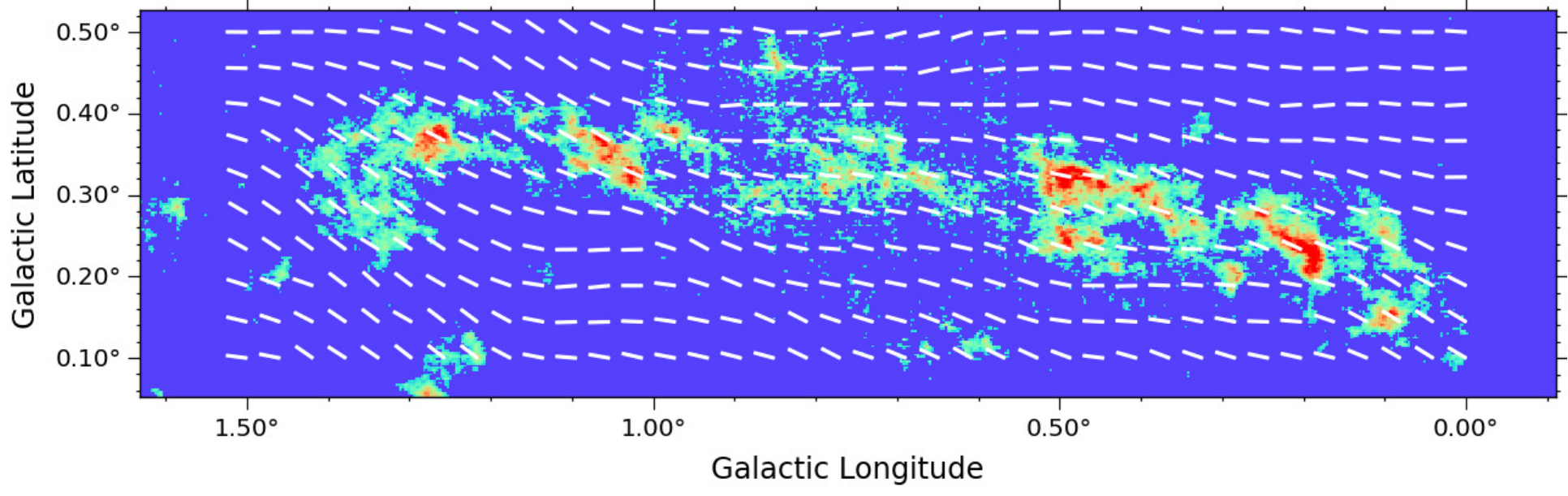}
\caption{$^{13}$CO integrated intensity map of the helix stream overlaid with Planck 353 GHz polarisation vectors (white). Note that the orientation of the magnetic field is perpendicular to the polarisation vectors.}
\label{planckpol}
\end{figure*}

\end{document}